\documentclass[aps,preprint,prd,showpacs,superscriptaddress,amssymb,amsmath,preprintnumbers,floatfix]{revtex4-1}
\usepackage{graphicx}
\usepackage{subfigure}
\usepackage{hyperref}
\usepackage{multirow}
\allowdisplaybreaks[1]
\usepackage[utf8]{inputenc}

\newcommand{\td}{{\rm{d}}}
\newcommand{\e}{{\rm{e}}}
\newcommand{\img}{{\rm{i}}}
\newcommand{\Tr}{{\rm{Tr}}}

\def\slashchar#1{\setbox0=\hbox{$#1$} 
\dimen0=\wd0 
\setbox1=\hbox{/} \dimen1=\wd1 
\ifdim\dimen0>\dimen1 
\rlap{\hbox to \dimen0{\hfil/\hfil}} 
#1 
\else 
\rlap{\hbox to \dimen1{\hfil$#1$\hfil}} 
/ 
\fi}

\begin{document}
\title{
$O(4)$-symmetric position-space renormalization of lattice operators
}

\newcommand{\Columbia}{
Physics Department,
Columbia University,
New York 10027, USA
}

\author{Masaaki~Tomii}
\email{mt3164_at_columbia.edu}
\author{Norman~H.~Christ}
\affiliation{\Columbia}



\begin{abstract}
We extend the position-space renormalization procedure, where
renormalization factors are calculated from Green's functions in
position space, by introducing a technique to take the average of
Green's functions over spheres.
In addition to reducing discretization errors, this technique enables
the resulting position-space correlators to be evaluated at any physical
distance, making them continuous functions similar to the
$O(4)$-symmetric position-space Green's functions
in the continuum theory but with a residual
dependence on a regularization parameter, the lattice spacing $a$.
We can then
take the continuum limit of these
renormalized quantities calculated at the same physical renormalization scale $|x|$
and investigate
the resulting $|x|$-dependence to identify the appropriate renormalization window.

As a numerical test of the spherical averaging technique,
we determine the renormalized light and strange quark masses by
renormalizing the scalar current.
We see a substantial reduction of discretization effects on the scalar
current correlator and an enhancement of the renormalization window.
The numerical simulation is carried out with $2+1$-flavor domain-wall
fermions at three lattice cutoffs in the range 1.79--3.15~GeV.
\end{abstract}

\maketitle

\section{Introduction}
\label{sec:intro}

Operator renormalization is necessary to calculate many quantities such as
weak matrix elements using lattice simulation.
So far, several different methods to renormalize lattice operators have been
proposed, applied and improved.
Since each method has individual advantages and disadvantages, we can
use the method that is the most convenient for our situation and purpose.

In this work, we focus on the position-space procedure~
\cite{Martinelli:1997zc,Gimenez:2004me}, 
in which the renormalization condition for an operator is imposed on the
corresponding correlation function in position space.
It is an important advantage of this procedure that it provides a fully
gauge invariant renormalization prescription since the correlator used in
the renormalization condition is gauge invariant.  
This advantage prevents the mixing with gauge noninvariant 
operators that occurs in gauge noninvariant schemes such as
the regularization independent momentum subtraction
(RI/MOM) scheme~\cite{Martinelli:1994ty}.  
Since the operators appearing in the position-space renormalization 
prescription are evaluated at separated space-time points, operators 
which vanish when the equations of motion are imposed will also not 
contribute.  Therefore a position-space renormalization scheme will
also avoid mixing with operators which vanish by the equations 
of motion -- mixing which can occur in the RI/MOM approach.

An important difficulty of the position-space approach arises from
the discrete lattice of points on which the position space Green's
function is evaluated.  Unless one works with lattices whose lattices
spacings are related as integer multiples, errors may be introduced
when combining results from two different ensembles.   Combining
results from ensembles with different lattice spacing is necessary 
both when evaluating the continuum limit and when using step scaling~
\cite{Luscher:1991wu,Jansen:1995ck,Arthur:2010ht}.  (For example,
when Cichy {\it et al.}~\cite{Cichy:2016qpu} employ step scaling
in position space they consider lattice spacings which differ by
factors of two.)   Recall that step scaling is an important nonperturbative 
method used to relate the normalization of operators that are being 
used in a coarse lattice calculation to physically equivalent operators 
defined on a fine, weak-coupling  lattice where a connection to 
perturbatively normalized operators can be more accurately made.  

In an RI/MOM scheme the Fourier transform averages over the
discrete lattice and the resulting functions of momentum approach
their continuum limits in a well-understood fashion~\cite{Symanzik:1983dc,
Symanzik:1983gh}.  In this paper we propose an alternative average that 
partially smooths the discrete nature of the position-space lattice while 
working with gauge invariant quantities and maintaining a non-zero 
separation between the operators whose Green's functions are being 
studied.

Our strategy is best illustrated using two-point functions, which are the
starting point of the present position-space renormalization schemes, as
illustrated by the expression
\begin{equation}
G(x_n) = \bigl\langle {\cal O}(x_n){\cal O}(0)^\dag \bigr\rangle
\label{eq: GF-2}
\end{equation}
where $\cal O$ is a gauge-invariant composite
local operator, $x_n$ is a point on our discrete lattice determined
by the four integers $n = (n_1, n_2, n_3, n_4)$ and, for 
simplicity, the second point $0$ is chosen to be the origin, also a 
point of this lattice.  As is described in more detail in 
Section~\ref{sec:shell_average}, we begin by extending this function 
into a function $\overline{G}(x)$ of the continuous position four-vector 
$x$, obtained by multi-linear interpolation from the sixteen values obtained 
by evaluating $G(x_n)$ at the sixteen lattice points that lie at the vertices 
of the four-dimensional cube which contains the point $x$.

Assuming, as we do throughout this paper, that our lattice theory has 
no order $a$ errors, our interpolated Green's function $\overline{G}(x)$
will agree with the corresponding Green's function of the continuum 
theory up to errors which vanish as $a^2$ in the continuum limit.
Of course, the $a^2$ errors which appear in the piece-wise linear 
function $\overline{G}(x)$ will still reflect the $O(4)$ symmetry 
breaking of the underlying lattice.  In order to reduce these lattice
artifacts and define a function of a single scale, we further simplify
our Green's function by averaging the point $x$ over a 
three-dimension sphere of radius $|x|$ centered at the origin:
\begin{equation}
\widehat{G}(|x|) =
\frac{1}{2\pi^2|x|^3}
\int \td^4 x'\, \delta\bigl(|x'|-|x|\bigr)\, \overline{G}(x').
\label{eq: GF-2-avg}
\end{equation}
We will then impose conditions on $\widehat{G}(|x|)$ to renormalize
the operator $O$.

The finite lattice spacing errors that are present in the lattice Green's 
function $G(x_n)$ are expected to appear as simple polynomials in
$a$ with coefficients which in perturbation theory depend only 
logarithmically in $a$, allowing a simple exptrapolation to the continuum
limit.  The lattice spacing dependence of our averaged quantity 
$\widehat{G}(|x|)$ will be more complicated.  In addition to the simple
$a^2$ errors coming from $G(x_n)$, the sphere averaging procedure
will introduce $O(a^2)$ errors which, while bounded by $a^2$ may be
complicated irregular functions of $a$ which could cause an explicit 
extrapolation in $a^2$ to fail.  As will be shown in 
Appendix~\ref{app:side_effect}, this irregular dependence on $a$ 
appears to be negligible, making the scheme proposed here suitable
for a calculation in which the continuum limit is to be evaluated.
Of course, were this error too large, we may be able to introduce a
higher-order interpolation scheme which would make these troubling 
effects of higher order than $a^2$ and therefore systematically
negligible.

As an example of the spherical average, we present our result for
the quark mass renormalization, which can be done by renormalizing
the scalar current.
There are several previous works on position-space renormalization
of bilinear operators~
\cite{Gimenez:2004me,Cichy:2012is,Tomii:2016xiv}.
For renormalization of bilinear operators, there is another important
advantage of the position-space procedure: the perturbative
matching to the modified minimal subtraction ($\rm\overline{MS}$) scheme
is available to $O(\alpha_s^4)$ for the vector, axial-vector, scalar and
pseudoscalar currents and to $O(\alpha_s^3)$ for the tensor current~
\cite{Chetyrkin:2010dx}.
Utilizing the spherical averaging technique, we perform a new analysis
that takes the continuum limit of the renormalized quark mass at
many values of $|x|$ and shows its $|x|$-dependence.
The final result agrees with the FLAG average~\cite{Aoki:2016frl}
as well as our previous result using the RI/SMOM scheme~
\cite{Aoki:2007xm,Sturm:2009kb}, an improved version of the RI/MOM scheme
with reduced sensitivity to long-distance effects, for the same ensembles~\cite{Blum:2014tka}.

An important future use for this sphere-averaged position-space 
renormalization scheme is to accurately define the weak operators
which are needed in the calculation of non-leptonic decays, such as
the $K\to\pi\pi$ decay, in a three-flavor theory.  At present these three-flavor 
operators are determined by using QCD perturbation theory to calculate that 
combination of three-flavor operators which will give the same
matrix elements as the more physical four-flavor operators when 
evaluated at energies below the charm threshold.  Such a use of
QCD perturbation theory below the charm threshold introduces 
uncontrolled systematic errors.  However, a nonperturbative
matching of three- and four-flavor operators using RI/MOM 
methods is also potentially uncertain.  The gauge-noninvariant
operators that are traditionally neglected in RI/MOM calculations 
when performed at higher energies because of the presence
of explicit factors of the gluon field, may give large contributions
at energies below the charm mass.  The sphere-averaged 
position-space renormalization scheme may allow a
nonperturbative determination of the three-flavor Wilson 
coefficients in which the only errors, which are systematically
improvable, come from the neglect of higher-dimension operators
proportional to inverse powers of the charm quark mass.

The paper is organized as follows.
In Section~\ref{sec:strategy}, we summarize the traditional
procedure of the position-space renormalization of an operator that
does not mix with any other operator and identify the problem posed
by the discretization errors that is addressed by the method
presented in this paper.
Our core technique in this work, the spherical average, is introduced
in Section~\ref{sec:shell_average}.
In Section~\ref{sec:renorm_qmass}, a concrete strategy to
calculate the renormalized quark mass through the position-space
renormalization of the scalar current is proposed.
In Section~\ref{sec:lattice_calculation}, the details of the numerical
simulation is described.
In Section~\ref{sec:result_mass}, our final result for the renormalized
quark mass is shown.
In the process, we show the performance of the
spherical average especially at short distances and discuss how the
renormalization window can be extended. 
In addition, we present a test of an {\it ad hoc} prescription to
reduce nonperturbative effects at long distances that are mainly due
to instanton interactions.
In Section~\ref{sec:conclusion}, we summarize the paper and
discuss the prospect of further applications of the spherical
average for various quantities calculated on the lattice.
In Appendix~\ref{app:side_effect}, we describe our investigation
of the irregular $a$-dependence that appears in the spherical
average, which turns out to be negligible.

\section{Fundamental procedure in previous works}
\label{sec:strategy}

In this section, we summarize the traditional approach to position-space
renormalization of an operator that does not mix with any other operator.
We consider two-point Green's functions of a composite operator
${\cal O}^s(\mu;x)$ renormalized at a scale $\mu$ in a scheme $s$ and the
corresponding lattice operator ${\cal O}^{\rm lat}(1/a;an)$ for a lattice
spacing $a$,
\begin{equation}
G_{\cal O}^s(\mu;x)
= \left\langle {\cal O}^s(\mu;x) {\cal O}^s(\mu;0)^\dag \right\rangle,
\ \ 
G_{\cal O}^{\rm lat}(1/a;an)
= \left\langle {\cal O}^{\rm lat}(1/a;an) {\cal O}^{\rm lat}(1/a;0)^\dag \right\rangle.
\label{eq:def_correlators}
\end{equation}
Here, we distinguish a four-dimensional point in the continuum theory
$x = (x_1,x_2,x_3,x_4)$ from that on the lattice
$an = (an_1,an_2,an_3,an_4)$ since the discrete character of the
lattice points is carefully considered throughout the paper.
In this section, we treat these two-point functions in the chiral limit,
which does not require consideration of the mass renormalization
of quarks in the correlators and remove an extra scale from the
renormalization procedure.

An operator ${\cal O}^X(\mu;x)$ renormalized at $\mu$ in the X-space
scheme \cite{Gimenez:2004me,Cichy:2012is} is defined in the continuum
theory by the condition
\begin{equation}
G_{\cal O}^X(\mu;x)\big|_{\mu=1/|x|} = G_{\cal O}^{\rm free}(x),
\label{eq:def_Xspace}
\end{equation}
where $G_{\cal O}^{\rm free}(x)$ is the corresponding two-point Green's
function evaluated in free field theory and $|x| = \sqrt{\sum_\mu x_\mu^2}$.
Since this nonperturbative scheme is fully gauge invariant and free
from contact terms unlike the RI/MOM scheme, it prevents mixing
with irrelevant operators and thus is a quite convenient scheme
especially at low energies where perturbative schemes are not applicable
and mixing with many irrelevant operators can occur in gauge
noninvariant schemes.

The traditional renormalization condition
\begin{equation}
{\widetilde Z_{\cal O}^{X/\rm lat}(\mu,1/a;an)}^2\big|_{\mu = 1/a|n|}
G_{\cal O}^{\rm lat}(1/a;an)
= G_{\cal O}^{\rm free}(x)\big|_{x=an},
\label{eq:renorm_cond0}
\end{equation}
yields 
\begin{equation}
\widetilde Z_{\cal O}^{X/\rm lat}(\mu,1/a;an)\big|_{\mu = 1/a|n|}
= \sqrt{
G_{\cal O}^{\rm free}(x)\big|_{x=an}
\over
G_{\cal O}^{\rm lat}(1/a;an)
},
\label{eq:ZtildeX}
\end{equation}
which violates rotational symmetry and depends on $n$ in a complicated way.
Since the $O(4)$-violating $n$-dependence is $O(a^2)$, it can be
eliminated and only the dependence on the distance scale $\mu = 1/a|n|$
remains if the continuum limit of the renormalized operator
${\widetilde Z_{\cal O}^{X/\rm lat}(\mu,1/a;an)}{\cal O}^{\rm lat}(1/a;an')$
is accurately taken.
However, evaluating the continuum limit requires an $a^2$
extrapolation of numerical values at
a fixed physical location $x = a_An_A = a_Bn_B=\ldots$ so that when
comparing ensembles $A$ and $B$ it is only the lattice spacing, not the
physical position which is changing.
This means the ratios of the lattice spacings for the ensembles
used to evaluate the continuum limit need to be integers or simple rational numbers.
However, lattice spacings are not tuned so precisely in practical simulations.
We propose a way to circumvent this problem in the next section.

We close the section by describing the relation between
operators in the $X$-space scheme and those in another scheme $s$.
Using Eqs.~\eqref{eq:def_correlators} and \eqref{eq:def_Xspace},
the matching factor $Z_{\cal O}^{s/X}(\mu,\mu')$, which is defined by
${\cal O}^s(\mu;x) = Z_{\cal O}^{s/X}(\mu,\mu'){\cal O}^X(\mu';x)$,
can be written as
\begin{equation}
Z_{\cal O}^{s/X}(\mu,\mu') =
\sqrt{
G_{\cal O}^s(\mu;x)
\over
G_{\cal O}^{\rm free}(x)
}\Bigg|_{|x| = 1/\mu'}.
\label{eq:matching}
\end{equation}
If we already know the correlator in the scheme $s$ and
any treatments in the X-space scheme such as the step scaling are not
needed, we can skip renormalizing operators to the X-space scheme and
directly compute 
\begin{equation}
\widetilde Z_{\cal O}^{s/\rm lat}(\mu,1/a;an)
\equiv Z_{\cal O}^{s/X}(\mu,\mu')
\widetilde Z_{\cal O}^{X/\rm lat}(\mu',1/a;an)
\big|_{\mu' = 1/a|n|}
= \sqrt{
G_{\cal O}^s(\mu;x)\big|_{x=an}
\over
G_{\cal O}^{\rm lat}(1/a;an)
}.
\label{eq:Ztilde}
\end{equation}
Of course, this expression violates rotational symmetry as does Eq.~\eqref{eq:ZtildeX}
and therefore
suffers from the same difficulty in taking the continuum limit of the corresponding
renormalized operator as is described above.

\section{Smoothing average over spheres}
\label{sec:shell_average}

Renormalization factors determined through the procedure
discussed in the previous section contain discretization errors which
depend in a complicated way on the lattice point $n$ where the renormalization
condition is imposed due to the violation of rotational symmetry. 
The complicated discretization errors induce difficulty in taking the continuum
limit of renormalized operators as mentioned in the previous section.
Some ideas to reduce this kind of discretization errors,
subtracting free-field discretization error~\cite{Gimenez:2004me,Tomii:2016xiv}
and discarding the lattice data points where discretization errors are quite large~
\cite{Chu:1993cn,Cichy:2012is}, have been applied.
These previous works usually na\"\i vely averaged the renormalization
factor Eq.~\eqref{eq:Ztilde} over lattice points in the renormalization window, which
could induce an irrelevant linear dependence on $a$ and further degrade the
accuracy of the continuum extrapolation of a renormalized quantity which
assumed that the leading discretization error is $O(a^2)$.
In this section, we propose another way to smooth lattice results, in which
the irrelevant $O(a^1)$ discretization error does not appear and the continuum
extrapolation of a renormalized quantity using a constant plus an $O(a^2)$
term can be safely taken.

We consider a lattice quantity $f_{a,n}$ calculated at each lattice point $n$.
The $a$-dependence of $f_{a,n}$ can be sketched as
\begin{equation}
f_{a,n} = F(x;a)|_{x=an} + c_{a,n}a^2 + O(a^4),
\label{eq:lattice_error}
\end{equation}
with a coefficient $c_{a,n}$ which depends on the lattice point $n$ in a complicated way.
In the simplest case, $F(x;a)$ is the continuum limit of the quantity being computed and
does not depend on $a$. However, by including a possible logarithmic $a$-dependence,
we can make our discussion more general and include the case where $f_{a,n}$ is an
$n$-dependent renormalization factor such as the quantities given in Eqs.~\eqref{eq:ZtildeX}
and \eqref{eq:Ztilde} or a correlator of unrenormalized operators.

We start with the case of one dimension, where we assume $x=x_1$.
We then use linear interpolation to extend the lattice results for $f_{a,n}$,
to define a function $\bar f_a(x)$ for all values of the continuous physical
distance $x$:
\begin{equation}
\bar f_a(x) = \frac{(a(n+1)-x)f_{a,n}+(x-an)f_{a,n+1}}{a},
\label{eq:eq_wave1dim}
\end{equation}
where $n$ is now defined as $\lfloor x/a\rfloor$, the largest integer that
is less than or equal to $x/a$.
Inserting Eq.~\eqref{eq:lattice_error} into this equation and expanding
$F(an;a)$ and $F(a(n+1);a)$
around $x$, we see that $\bar f_a(x)$ is an approximation to $F(x;a)$
as a continuous function of $x$ that is accurate up to $O(a^2)$.
Note that the appropriate weight of $a(n+1)-x$ and $x-an$ in
Eq.~\eqref{eq:eq_wave1dim} is important to avoid introducing an $O(a^1)$ error which
would spoil the accuracy of an $a^2$ continuum extrapolation.

In the case of two dimensions, the weighted average Eq.~\eqref{eq:eq_wave1dim}
can be modified to a bilinear interpolation
\begin{align}
\bar f_a(x)
&= a^{-2}(
\begin{array}{cc}
a(n_1+1)-x_1 & x_1-an_1
\end{array}
)\left(
\begin{array}{cc}
f_{a,n} & f_{a,n+\hat 2}
\\
f_{a,n+\hat 1} & f_{a,n+\hat 1+\hat 2}
\end{array}
\right)\left(
\begin{array}{c}
a(n_2+1)-x_2 \\ x_2-an_2
\end{array}
\right),
\label{eq:bilin_interpolate}
\end{align}
where $n_\mu = \lfloor x_\mu/a\rfloor$ and $\hat\mu$ is the unit vector for
the $\mu$-direction.
While this weighted average is also easily found to be free from the $O(a^1)$ error,
it is expected to depend significantly on the
direction of $x$ as well as the distance $|x|$ due to the violation of rotational
symmetry.
The most na\"\i ve way to smooth this discretization error may be to
introduce the average over a circle with the radius of $|x|$,
\begin{equation}
\hat f_a(|x|)
= {1\over2\pi}\int_0^{2\pi}\td\theta\,\bar f_a(x),
\label{eq:sphe_ave2}
\end{equation}
where we use two-dimensional polar coordinates
\begin{equation}
x_1 = |x|\cos\theta,\ \ x_2 = |x|\sin\theta.
\end{equation}

The extension to four dimensions is straightforward.
The interpolation of $f_a$ at $x$ is given by
\begin{equation}
\bar f_a(x) = a^{-4}\sum_{i,j,k,l=0}^1\Delta_{1,i}\Delta_{2,j}\Delta_{3,k}\Delta_{4,l}\,
f_{a,n+i\hat1+j\hat2+k\hat3+l\hat4},
\label{eq:quadrilin_interpolate}
\end{equation}
where we define the factors
\begin{equation}
\Delta_{\mu,i} = |a(n_\mu+1-i)-x_\mu|.
\end{equation}
One can easily verify this interpolated value is also free from the $O(a^1)$ error.
The smoothing average over the four-dimensional sphere with the radius of $|x|$ is
\begin{equation}
\hat f_a(|x|)
={1\over2\pi^2}\int_0^\pi\td\theta_1\int_0^\pi\td\theta_2\int_0^{2\pi}\td\theta_3
\sin^2\theta_1\sin\theta_2\,\bar f_a(x),
\label{eq:sphe_ave4}
\end{equation}
with four-dimensional polar coordinates
\begin{align}
x_1 &= |x|\cos\theta_1,
\notag\\
x_2 &= |x|\sin\theta_1\cos\theta_2,
\notag\\
x_3 &= |x|\sin\theta_1\sin\theta_2\cos\theta_3,
\notag\\
x_4 &= |x|\sin\theta_1\sin\theta_2\sin\theta_3.
\end{align}
The averaged quantity $\hat f_a(|x|)$ will differ from the direction independent
continuum quantity $F(x;a)$ by discretization errors of $O(a^2)$.

Although the discretization error of the spherical average is thus $O(a^2)$,
it should be noted that the averaged value is not a regular polynomial
in $a$ but will contain extra non-differentiable terms of $O(a^2)$
because of the complicated $a$-dependence of the floor function
$n_\mu = \lfloor x_\mu/a\rfloor$.
This irregularity could arise also from the fact that the set of the lattice points
$n$ and their weight used by the spherical average at each fixed physical
distance $|x|$ depend on the lattice spacing $a$.
Such complicated $a$-dependence could spoil the accuracy of a continuum
extrapolation which assumed a regular $a^2$ term.
In Appendix~\ref{app:side_effect}, we discuss the significance
of such complicated $a$-dependence and demonstrate it is small.

\section{Quark masses renormalization in position space}
\label{sec:renorm_qmass}

\subsection{Strategy}
\label{sec:mass_renorm}

Since the quark mass renormalization factor $Z_m$ can be calculated as the
inverse of the renormalization factor $Z_S$ of the scalar current $S(x) = \bar u(x)d(x)$,
we consider the renormalization of $S(x)$, which is equivalent to that of the
pseudoscalar current $P(x) = \bar u(x)\img\gamma_5d(x)$ as long as chiral
symmetry on the lattice is maintained.
Since we use domain-wall fermions, we can calculate $Z_m$ from the
renormalization of $S(x)$ and $P(x)$.

In what follows, we employ the $\rm\overline{MS}$ scheme and
introduce the input light quark mass parameter $m_{ud}'$ that is
used for the calculation of the correlators on the lattice.
The $n$-dependent renormalization factor Eq.~\eqref{eq:Ztilde} is then
rewritten as
\begin{equation}
\widetilde Z^{\rm\overline{MS}/lat}_{S/P} (\mu,1/a;an;m_{ud}')
= \sqrt{\frac{G_S^{\rm\overline{MS}}(\mu;x;0)\big|_{x=an}}{G_{S/P}^{\rm lat}(1/a;an;m_{ud}')}}.
\label{eq:renorm_cond}
\end{equation}
The chiral limit ($m_{ud}'\to0$) is taken in Section~\ref{sec:result_mass}.
In this work, 
the scalar correlator $G_S^{\rm\overline{MS}}$
in continuum perturbation theory is considered only in the massless limit,
where it is equivalent to the pseudoscalar correlator
and is available to $O((\alpha_s/\pi)^4)$ accuracy~\cite{Chetyrkin:2010dx}.
The strategy to improve the convergence of the perturbative series of the correlator
is discussed in the following subsection and in~\cite{Tomii:2016xiv} for more detail.

We also analyze an $O(4)$-symmetric renormalization factor
\begin{equation}
\widehat{\widetilde Z}_{S/P}^{\rm\overline{MS}/lat}(\mu,1/a;|x|;m_{ud}')
= Z_S^{{\rm\overline{MS}}/X}(\mu,\mu')
\widehat{\widetilde Z}_{S/P}^{X/\rm lat}(\mu',1/a;|x|;m_{ud}')
\label{eq:ZsO4MSbar}
\end{equation}
obtained from Eq.~\eqref{eq:matching} and the $O(4)$-symmetric
renormalization condition
\begin{equation}
\widehat{\widetilde Z}_{S/P}^{X/\rm lat}(\mu,1/a;|x|;m_{ud}')^2\big|_{\mu=1/|x|}
\widehat G_{S/P}^{\rm lat}(1/a;|x|) = G_S^{\rm free}(x),
\label{eq:O4condition}
\end{equation}
with the sphere-averaged Green's function $\widehat G_S^{\rm lat}(1/a;|x|)$
calculated as follows.
It should be noted that the complicated $a$-dependence appearing in the
multi-linear interpolation depends on the first and second derivatives of
the continuum version of the function that is to be interpolated with
respect to $|x|$.
Therefore, the spherical averaging procedure is applied to a function whose
$|x|$-dependence in the continuum limit is as small as possible.
For this reason, we calculate the spherical average of the ratio
$G_{S/P}^{\rm lat}(1/a;an;m_{ud}')/G_S^{\rm free}(x)|_{x=an}$
at each distance $|x|$ and then define the sphere-averaged Green's
function $\widehat G_S^{\rm lat}(1/a;|x|)$ as the product of it and
$G_S^{\rm free}(x)$.

Note that either $\widetilde Z^{\rm\overline{MS}/lat}_{S/P} (\mu,1/a;an;m_{ud}')$ 
or $\widehat{\widetilde Z}_{S/P}^{\rm\overline{MS}/lat}(\mu,1/a;|x|;m_{ud}')$
may not be an appropriate renormalization factor since it still depends on the location
$n$ or $|x|$ due to the following sources of error:
\begin{itemize}
\item Discretization effects in $G_{S/P}^{\rm lat}(1/a;an;m_{ud}')$.
\item Truncation error from the perturbative calculation of $G_S^{\rm\overline{MS}}(\mu;x;0)$.
\item Nonperturbative QCD effects, which are not present in the
perturbatively calculated $G_S^{\rm\overline{MS}}(\mu;x;0)$
but do appear in the nonperturbatively measured $G_{S/P}^{\rm lat}(1/a;an;m_{ud}')$.
\end{itemize}
The first source is uncontrollable at short distances ($|x|,a|n|\sim a$), while the others
are significant at long distances ($|x|, a|n|\gtrsim1/\Lambda_{\rm QCD}$).
We need to find or create an appropriate window where all of these sources
of error are under control and the $n$-dependence of
$\widetilde Z^{\rm\overline{MS}/lat}_{S/P} (\mu,1/a;an;m_{ud}')$ 
or $|x|$-dependence of 
$\widehat{\widetilde Z}_{S/P}^{\rm\overline{MS}/lat}(\mu,1/a;|x|;m_{ud}')$
is sufficiently small.
Since the third source especially violates the degeneracy of
$\widetilde Z^{\rm\overline{MS}/lat}_S (\mu,1/a;an;m_{ud}')$ and
$\widetilde Z^{\rm\overline{MS}/lat}_P (\mu,1/a;an;m_{ud}')$,
analyzing both of these may specify the region where nonperturbative
effects are less significant.

Using the unrenormalized quark mass $m_q^{\rm bare}(1/a)$ at the physical
pion mass, which is given in Ref.~\cite{Blum:2014tka} for the degenerate up and
down quarks  $(q=ud)$ and the strange quark $(q=s)$ on our ensembles,
we analyze the $n$- and $|x|$-dependent renormalized quark masses
\begin{equation}
\widetilde m_{q,S/P}^{\rm\overline{MS}}(\mu;an;a,m_{ud}')
= \frac{m_q^{\rm bare}(1/a)}{\widetilde Z_{S/P}^{\rm\overline{MS}/lat}(\mu,1/a;an;m_{ud}')},
\label{eq:eq:renorm_mass}
\end{equation}
and
\begin{equation}
\widehat{\widetilde m}_{q,S/P}^{\rm\overline{MS}}(\mu;|x|;a,m_{ud}')
= \frac{m_q^{\rm bare}(1/a)}{\widehat{\widetilde Z}_{S/P}^{\rm\overline{MS}/lat}(\mu,1/a;|x|;m_{ud}')},
\label{eq:eq:renorm_mass_ave}
\end{equation}
where $q = ud, s$.

In Section~\ref{sec:result_mass}, we determine the renormalized mass of the
degenerate up and down quarks and the strange quark on our ensembles.

\subsection{Scalar correlator in massless perturbation theory}
\label{sec:MLPT}

While the available four-loop perturbative results is an important advantage of the
position-space renormalization of the scalar current,
the region where discretization errors may be
under controlled is $1/|x| \lesssim1$~GeV for currently available lattices
with domain-wall fermions and therefore the convergence of
the perturbative expansion might be still insufficient.
The convergence can be improved by a resummation of the perturbative series
using the coupling constant at another renormalization scale as explained below.

Chetyrkin and Maier~\cite{Chetyrkin:2010dx} gave the coefficients $C_i^{S,\rm CM}$
of the perturbative expansion
\begin{equation}
G_S^{\rm\overline{MS}}(\tilde\mu_x;x;0)
= \frac{3}{\pi^4|x|^6}\left(1+\sum_iC_i^{S,\rm CM}a_s(\tilde\mu_x)^i\right),
\label{eq:PT_CM}
\end{equation}
up to $i=4$.
Here, the strong coupling constant $a_s(\tilde\mu_x) = \alpha_s(\tilde\mu_x)/\pi$ is
renormalized in the $\rm\overline{MS}$ scheme at
$\tilde\mu_x = 2\e^{-\gamma_E}/|x|\simeq1.123/|x|$ with Euler's constant
$\gamma_E = 0.5772$ and is evaluated using the scale of QCD
$\Lambda_{\rm QCD}^{\rm\overline{MS}} = 332(17)$~\cite{Tanabashi:2018oca}
in three flavor theory.
By setting the renormalization scale $\tilde\mu_x$ of the scalar current
and the strong coupling constant proportional to $|x|^{-1}$, the
logarithmic $|x|$-dependence of the perturbative coefficients can be eliminated.

The anomalous dimension of the scalar current, which is the same as the
mass anomalous dimension except for the sign and is calculated up to the
five-loop level~\cite{Baikov:2014qja}, enables us to evolve the scale on the
LHS of Eq.~\eqref{eq:PT_CM}.
The beta function, which is also available to the five-loop level~\cite{Baikov:2016tgj},
can be used to evolve the scale of the strong coupling constant on the RHS of
Eq.~\eqref{eq:PT_CM}.
Using the original perturbative coefficients $C_i^{S,\rm CM}$ and these
scale evolution procedures, we obtain a general expression of the perturbative
series
\begin{equation}
G_S^{\rm\overline{MS}}(\mu_x';x;0)
= \frac{3}{\pi^4x^6}\left(1+\sum_iC_i^S(\mu_x^*,\mu_x')a_s(\mu_x^*)^i\right),
\end{equation}
where $\mu_x'$ and $\mu_x^*$ are the renormalization scale of the scalar current
and that of the strong coupling constant, respectively.
While the all-order calculation of the RHS is supposed to be independent of $\mu_x^*$,
any finite-order calculation does depend on $\mu_x^*$.
Therefore, the convergence of the perturbative series can be investigated by varying
$\mu_x^*$.

Thus, we obtain the numerical value of the scalar correlator
$G_S^{\rm\overline{MS}}(\mu_x';x;0)\big|_{\mu_x^*}$ calculated with a scale
$\mu_x^*$ of the strong coupling constant.
In order to renormalize the scalar current at a specific scale, which we set to 3~GeV,
the scale evolution is needed from $\mu_x'$ to 3~GeV,
\begin{align}
G_S^{\rm\overline{MS}}({\rm 3\,GeV};x;0)\big|_{\mu_x^*,\mu'_x}
&= \exp\left(
-2\int_{a_s(\mu_x')}^{a_s({\rm3\,GeV})}\frac{\td z}{z}\frac{\gamma_m(z)}{\beta(z)}
\right)G_S^{\rm\overline{MS}}(\mu_x';x;0)\big|_{\mu_x^*}
\notag\\
&= \left(\frac{\rho(a_s(\mu_x'))}{\rho(a_s({\rm3\,GeV}))}\right)^2
G_S^{\rm\overline{MS}}(\mu_x';x;0)\big|_{\mu_x^*},
\end{align}
where $\gamma_m(z)$ and $\beta(z)$ are the mass anomalous dimension
and the beta function, respectively, and $\rho(z)$ is known to the five-loop level~
\cite{Baikov:2014qja}.
While $G_S^{\rm\overline{MS}}({\rm 3\,GeV};x;0)\big|_{\mu_x^*,\mu'_x}$
is also supposed to be independent of $\mu_x^*$ and $\mu_x'$ in an
all-order calculation, the convergence can be optimized by tuning the scale
parameters $\mu_x'$ and $\mu_x^*$ so that the dependence on
these scale parameters is minimized.
We use the optimal values $\mu_x' = \e^{0.8}/|x| \simeq2.2/|x|$ and
$\mu_x^* = \e^{1.05}/|x| \simeq2.9/|x|$ in the case of three-flavor QCD
quoted by Ref.~\cite{Tomii:2016xiv}.

\section{Lattice setup}
\label{sec:lattice_calculation}

\tabcolsep = 6pt
\begin{table}[htb]
\caption{
Lattice ensembles used in this work.
}
\label{tab:ensembles}
\begin{center}
\begin{tabular}{cccccccc}
\hline
\hline
Ensemble set & $\beta$ & $a^{-1}$ [GeV] & $L^3\times T\times L_s$
& $am_s'$ & $am_{ud}'$ & $aM_\pi$ & $N_{\rm conf}$\\
\hline
24I & 2.13 & 1.785(5) & $24^3\times64\times16$ & 0.0400 & 0.0050 & 0.1904(6) & 137\\ 
& & & & & 0.0100 & 0.2422(5) & 77\\ 
\hline
32I & 2.25 & 2.383(9) & $32^3\times64\times16$ & 0.0300 & 0.0040 & 0.1269(4) & 157\\ 
& & & & & 0.0080 & 0.1727(4) & 110\\ 
\hline
32Ifine & 2.37 & 3.148(17) & $32^3\times64\times12$ & 0.0186 & 0.0047 & 0.1179(13) & 170\\ 
\hline
\hline
\end{tabular}
\end{center}
\end{table}

We perform lattice simulation with the ensembles of $2+1$-flavor dynamical
domain-wall fermions~\cite{Kaplan:1992bt,Shamir:1993zy} and the Iwasaki
gauge action~\cite{Iwasaki:1984cj,Iwasaki:1985we} generated by the RBC
and UKQCD collaborations~\cite{Blum:2014tka}.
Table~\ref{tab:ensembles} summarizes the properties of the ensembles used
in this work. 
We calculate with three lattice cutoffs ranging from 1.785(5)~GeV to
3.148(17)~GeV.
The coarsest ensembles (24I) and the finer ones (32I and 32Ifine) 
are generated on the $24^3\times64$ and $32^3\times64$ lattices,
respectively.
The strange quark mass $m_s'$ is used only for the strange sea quark, while
we use the same values of the sea and valence quark masses $m_{ud}'$
for the degenerate up and down quarks.
The corresponding pion masses of the ensembles are quoted from
Refs.~\cite{Aoki:2010dy,Blum:2014tka} and in the region from 300~MeV to
430~MeV.

\tabcolsep = 6pt
\begin{table}[tbp]
\caption{
Values of $Z_q(1/a)$ quoted from Ref.~\cite{Blum:2014tka}.
}
\label{tab:Zq}
\begin{center}
\begin{tabular}{ccc}
\hline
\hline
Ensemble set & $Z_{ud}$ & $Z_s$\\
\hline
24I & 0.9715(54) & 0.9628(40) \\ 
32I & 1 & 1 \\
32Ifine & 1.015(17) & 1.005(12) \\
\hline
\hline
\end{tabular}
\end{center}
\end{table}

We distinguish the input quark masses $m_q'$ $(q = ud,s)$, which are used in
the lattice calculations and shown in Table~\ref{tab:ensembles},
from the renormalized and unrenormalized quark masses
($m_q^{\rm\overline{MS}}(\mu)$ and $m_q^{\rm bare}(1/a)$) that
realize the physical pion mass.
In Ref.~\cite{Blum:2014tka}, the values of unrenormalized quark masses
were represented by
\begin{equation}
m_q^{\rm bare}(1/a) = \frac{m_q^{\rm bare,32I}}{Z_q(1/a)},
\label{eq:def_mqbare}
\end{equation}
where the values of quantities on the RHS were obtained by a global
continuum and chiral fit to ten ensembles in the continuum scaling with
the input experimental values of pion, kaon and Omega baryon masses
\cite{Blum:2014tka}.
We use
$m_{ud}^{\rm bare,32I} = 2.198(11)$~MeV,
$m_{ud}^{\rm bare,32I} = 60.62(24)$~MeV and 
$Z_q(1/a)$ summarized in Table~\ref{tab:Zq}.

For each configuration, we calculate the scalar and pseudoscalar
correlators with 16 point sources located at
\begin{equation}
\begin{array}{c}
(0,0,0,0),\ \ (\frac{L}{2},\frac{L}{2},0,0),\ \ (0,\frac{L}{2},\frac{L}{2},0),\ \ (\frac{L}{2},0,\frac{L}{2},0),\\
(\frac{L}{4},\frac{L}{4},\frac{L}{4},\frac{T}{4}),\ \ (\frac{3L}{4},\frac{3L}{4},\frac{L}{4},\frac{T}{4}),\ \ (\frac{L}{4},\frac{3L}{4},\frac{3L}{4},\frac{T}{4}),\ \ (\frac{3L}{4},\frac{L}{4},\frac{3L}{4},\frac{T}{4}),\\
(\frac{L}{2},\frac{L}{2},\frac{L}{2},\frac{T}{2}),\ \ (0,0,\frac{L}{2},\frac{T}{2}),\ \ (\frac{L}{2},0,0,\frac{T}{2}),\ \ (0,\frac{L}{2},0,\frac{T}{2}),\\
(\frac{3L}{4},\frac{3L}{4},\frac{3L}{4},\frac{3T}{4}),\ \ (\frac{L}{4},\frac{L}{4},\frac{3L}{4},\frac{3T}{4}),\ \ (\frac{3L}{4},\frac{L}{4},\frac{L}{4},\frac{3T}{4}),\ \ (\frac{L}{4},\frac{3L}{4},\frac{L}{4},\frac{3T}{4}),
\end{array}
\end{equation}
and average the correlators over all these source points.

\section{Numerical results}
\label{sec:result_mass}

\begin{figure}[tbp]
\begin{center}
\subfigure{\mbox{\raisebox{1mm}{\includegraphics[width=110mm, bb=0 0 345 230]{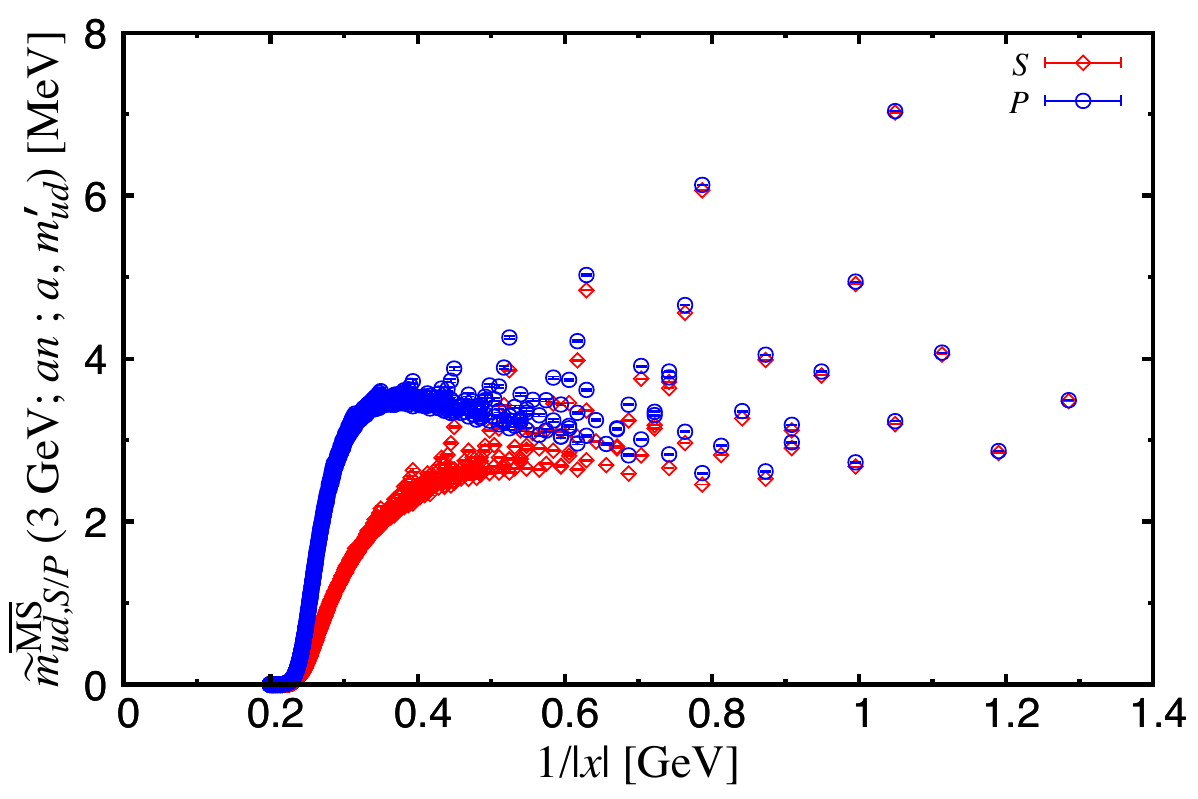}}}}
\caption{
Results for $\widetilde m_{ud,S/P}^{\rm\overline{MS}}(3~{\rm GeV};an;a,m_{ud}')$ calculated on
the 32Ifine ensemble.
The perturbative calculation is done at $\mu_x^* = \e^{1.05}/|x|$ and $\mu_x' = \e^{0.80}/|x|$.
}
\label{fig:mudMSb3GeV_lat_indiv}
\end{center}
\end{figure}

Figure~\ref{fig:mudMSb3GeV_lat_indiv} shows
$\widetilde m_{ud,S/P}^{\rm\overline{MS}}(3~{\rm GeV};x;a,m_{ud}')$ defined in
Eq.~\eqref{eq:eq:renorm_mass} calculated on the 32Ifine ensemble.
The results for both the scalar (diamonds) and pseudoscalar (circles) channels
are shown.
Because of the violation of rotational symmetry, we distinguish, in the
figure, different lattice points that are not equivalent with respect to $90^\circ$
rotations or parity inversion in the four-dimensional hypercubic group.
For example, (1,1,1,1) and (2,0,0,0) correspond to the same distance
$a|n|$ but are distinguished since they are indeed different points if
rotational symmetry is violated and only hypercubic symmetry remains.
The results are averaged over sets of lattice points related by
$90^\circ$ rotations including parity inversion.

The $n$-dependence of this quantity arises mainly from discretization
errors at short distances, the truncation error of the
perturbative calculation and nonperturbative effects
at long distances as explained in Section~\ref{sec:mass_renorm}.
These sources of the $n$-dependence need to be
under controlled in order to obtain the correct value of the renormalized
mass $m_q^{\rm\overline{MS}}(3~\rm GeV)$.
However, Figure~\ref{fig:mudMSb3GeV_lat_indiv} indicates the ambiguity due
to such $n$-dependence is $O(1~{\rm MeV})$, which is much larger than the
uncertainty of the renormalized light quark mass calculated by other works.

A rapid decrease is seen below $1/|x|=1/a|n| \sim0.3$~GeV 
since the truncation uncertainty of the perturbative calculation
increases tremendously below this threshold.
We do not expect that this lower limit on the
perturbative window can be decreased because
the convergence is already optimized by our choice of
$\mu_x^* = \e^{1.05}/|x|$ and $\mu_x' = \e^{0.80}/|x|$.
These choices are found to maintain reasonable convergence
down to $1/|x|\sim0.4$~GeV~\cite{Tomii:2016xiv}.

Among the three sources of $n$-dependence listed in
Section~\ref{sec:mass_renorm}, the $n$-dependence associated
with the convergence of the perturbative calculation
is thus already taken into account as much as possible.
We discuss and take into account the remaining two sources below.
The $n$-dependence associated with discretization errors can be
reduced by the spherical average designed in Section~\ref{sec:shell_average}.
The third source of $n$-dependence associated with nonperturbative effects
can be investigated by comparing the scalar and pseudoscalar channels.
While Figure~\ref{fig:mudMSb3GeV_lat_indiv} provides some
information, we prefer to take the spherical average first and then to
discuss the difference between the scalar and pseudoscalar correlators.

\begin{figure}[tbp]
\begin{center}
\subfigure{\mbox{\raisebox{1mm}{\includegraphics[width=110mm, bb=0 0 345 230]{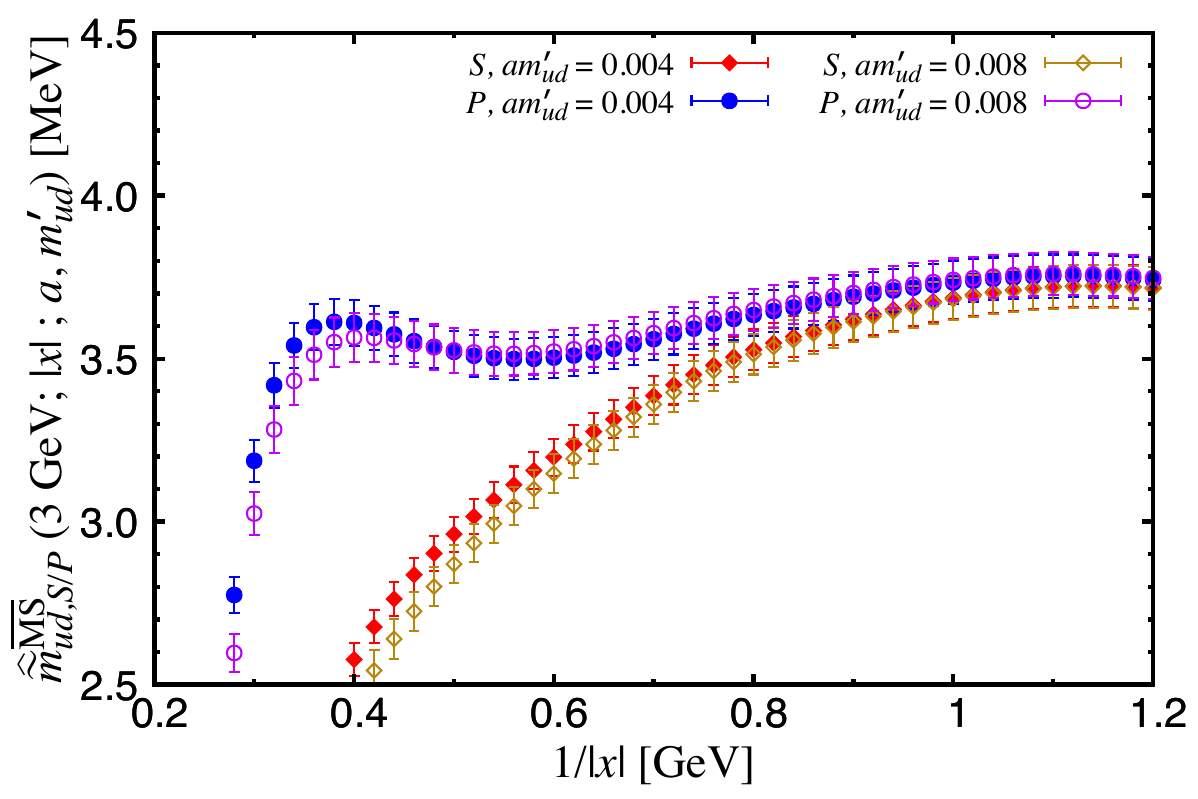}}}}
\caption{
Results for 
$\widehat{\widetilde m}_{ud,S/P}^{\rm\overline{MS}}(3~{\rm GeV};|x|;a,m_{ud}')$
calculated on the 32I ensembles.
}
\label{fig:mudMSb3GeV_ps32I}
\end{center}
\end{figure}

Figure~\ref{fig:mudMSb3GeV_ps32I} shows the result for
$\widehat{\widetilde m}_{ud,S/P}^{\rm\overline{MS}}(3~{\rm GeV};x;a)$
defined in Eq.~\eqref{eq:eq:renorm_mass_ave}, where the renormalization
factors of the scalar (diamonds) and pseudoscalar (circles) currents are
calculated using the spherical average  of the corresponding Green's
functions calculated on the 32I ensembles with $am_{ud}' = 0.004$
(filled points) and $am_{ud}' = 0.008$ (open points).
While discretization errors appear to be much reduced and the
the $n$-dependence is made smaller, there must still be
some dependence on the regularization parameter $a$ in the form
of $a^2/x^2$, $a^2\Lambda_{\rm QCD}^2$, $a^4/|x|^4$ and so on.
Therefore, the continuum extrapolation should be performed at sufficiently
long distances where only $O(a^2)$ discretization errors are visible.
On the other hand, the difference between the scalar and pseudoscalar
channels is significant, 1\% at $1/|x| \simeq 1$~GeV and 3\% at
$1/|x| \simeq 0.8$~GeV, although we use domain-wall fermions which have
high degree of chiral symmetry.
Therefore, the extraction of the quark mass should be done at sufficiently
short distances, $1/|x| \gtrsim 1$~GeV in order to obtain
a systematic error less than 1\%.
For this, the continuum extrapolation has to be safe at $1/|x| \simeq 1$~GeV.

An important advantage of the spherical average is our ability to
take the continuum limit of renormalized quantities as explained below.
Since the structure of the $a$-dependence depends on $|x|$ as it contains a
term proportional to $a^2/x^2$, the renormalized quantities need to be
calculated at the same physical distance scale $|x|$ for each ensemble in
order to take the continuum limit as a quadratic $a^2\to0$ extrapolation.
The spherical averaging technique trivially enables such an extrapolation.
The extrapolation of the spherical average
$\widehat{\widetilde m}_{ud,S/P}^{\rm\overline{MS}}(3~{\rm GeV};|x|;a,m_{ud}')$
to the continuum ($a\to0$) and chiral ($m_{ud}'\to0$) limits is done by
performing a simultaneous fit to the data from all the ensembles with the
fit function
\begin{equation}
\widehat{\widetilde m}_{ud,S/P}^{\rm\overline{MS}}(3~{\rm GeV};|x|;a,m_{ud}')
= \widehat{\widetilde m}_{ud,S/P}^{\rm\overline{MS}}(3~{\rm GeV};|x|)
+ C_{a,S/P}(|x|)a^2 + C_{m,S/P}(|x|)M_\pi(a,m_{ud}')^2,
\label{eq:extrapolation}
\end{equation}
with three fit parameters: $\widehat{\widetilde m}_{ud,S/P}^{\rm\overline{MS}}(3~{\rm GeV};|x|)$,
$C_{a,S/P}(|x|)$ and $C_{m,S/P}(|x|)$ for each $|x|$.
Here, we introduce a term proportional to the pion mass squared
$M_\pi(a,m_{ud}')^2$ labeled by the ensemble parameters $a$ and $m_{ud}'$,
although the leading mass correction in perturbation theory is proportional to
quark mass squared or $M_\pi(a,m_{ud}')^4$.
This is because the perturbative mass correction is much smaller than
the mass correction from OPE such as $m\langle \bar qq\rangle$
and $m\langle \bar qGq\rangle$ around $1/|x|\sim0.5$~GeV
\cite{Chu:1993cn,Hands:1994cj}.

\begin{figure}[tbp]
\begin{center}
\subfigure{\mbox{\raisebox{1mm}{\includegraphics[width=110mm, bb=0 0 345 230]{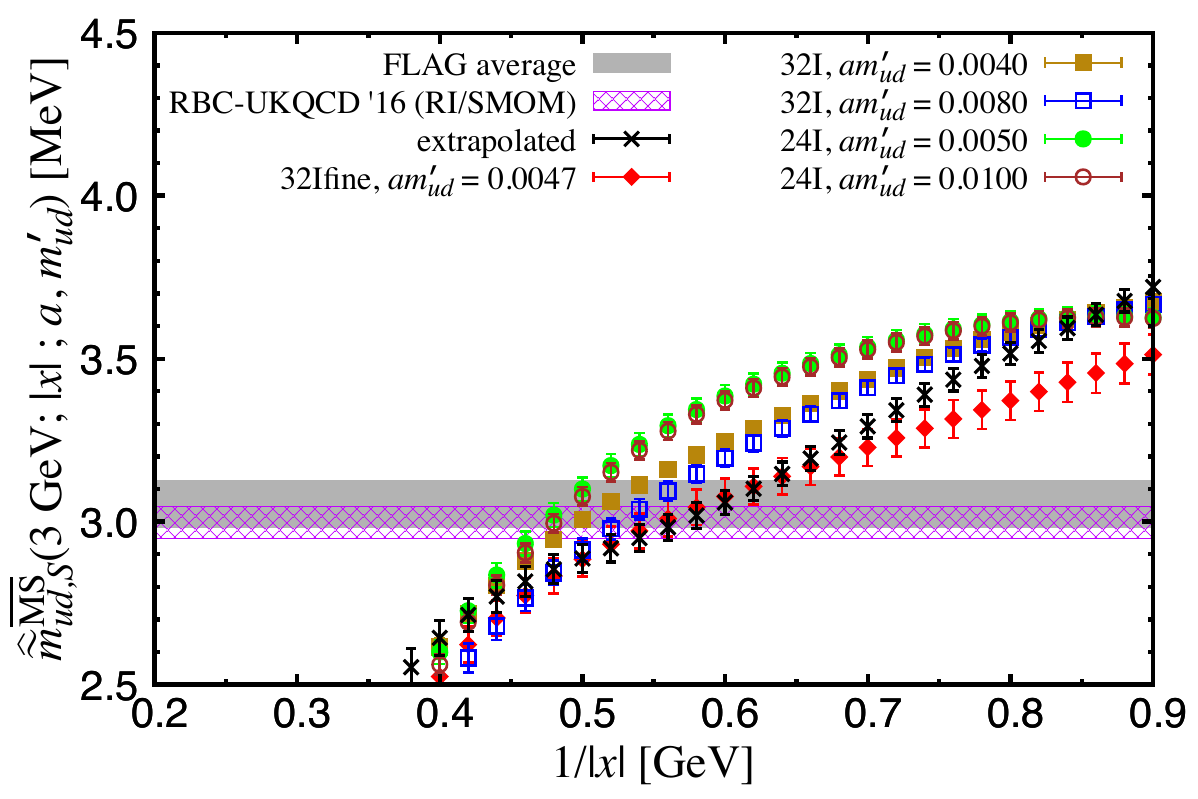}}}}
\subfigure{\mbox{\raisebox{1mm}{\includegraphics[width=110mm, bb=0 0 345 230]{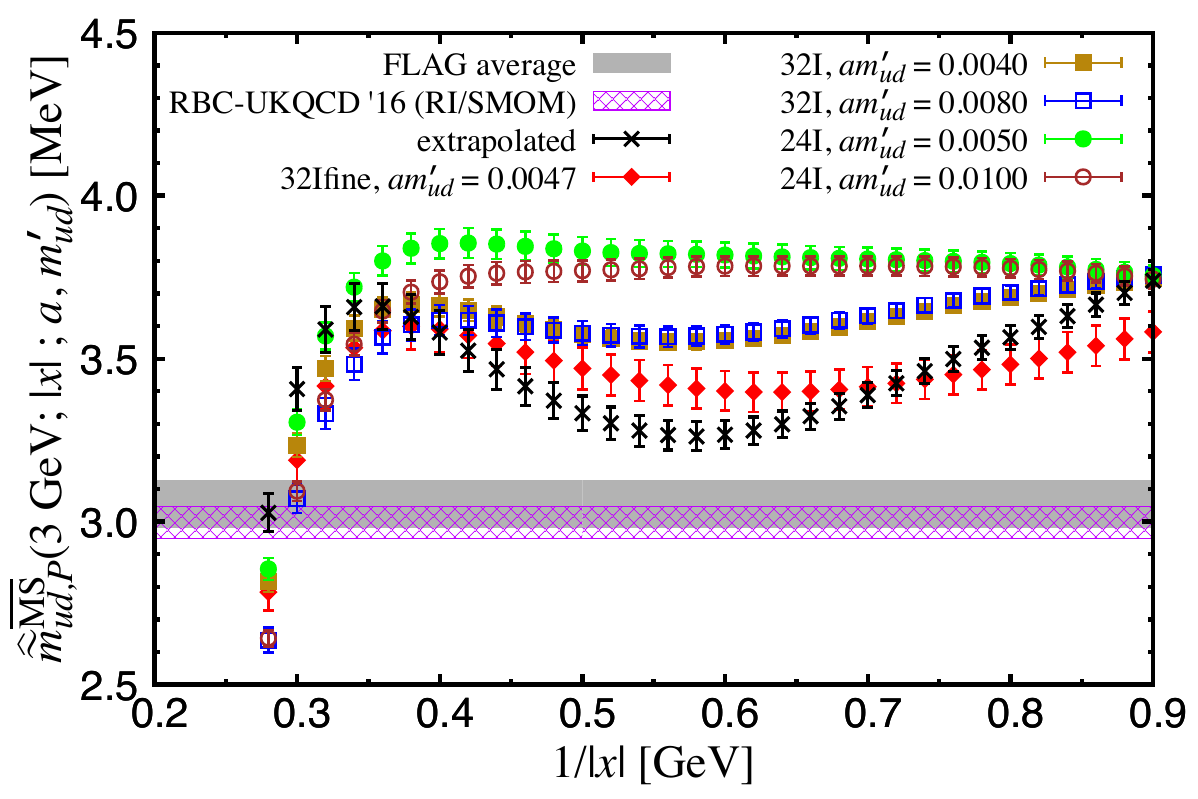}}}}
\caption{
Results for the spherical average
$\widehat{\widetilde m}_{ud,S/P}^{\rm\overline{MS}}(3~{\rm GeV};|x|;a,m_{ud}')$
calculated on all the ensembles listed in Table~\ref{tab:ensembles}
and its extrapolation to the continuum and chiral limits ($a\to0, m_{ud}'\to0$).
The results for the scalar (upper panel) and pseudoscalar (lower panel) channels
are shown separately.
}
\label{fig:mudMSb3GeV_ps}
\end{center}
\end{figure}

Figure~\ref{fig:mudMSb3GeV_ps} shows the result for the spherical average
$\widehat{\widetilde m}_{ud,S/P}^{\rm\overline{MS}}(3~{\rm GeV};|x|;am_{ud}')$
on each ensemble listed in Table~\ref{tab:ensembles} and its continuum and
chiral limits
$\widehat{\widetilde m}_{ud,S/P}^{\rm\overline{MS}}(3~{\rm GeV};|x|)$.
We also show the FLAG average~\cite{Aoki:2016frl} (solid band) and our
previous RBC/UKQCD result obtained through the RI/SMOM scheme~\cite{Blum:2014tka}
(hatched band), which is $m_{ud}^{\rm\overline{MS}}(3~{\rm GeV}) = 2.997(49)$~MeV
including the statistical and systematic errors.
While FLAG gave the value renormalized at 2~GeV~\cite{Aoki:2016frl},
we perform its scale evolution to 3~GeV and show
$m_{ud}^{\rm\overline{MS}}(3~{\rm GeV}) = 3.054(72)$~MeV
in the figure.
We show the result only for $1/|x|\le0.9$~GeV because the spherical average
with $|x|<2a$, which corresponds to $1/|x|>0.89$~GeV on the coarsest lattice,
uses the value at $x=0$ and therefore contains unphysical contact terms.
Since there is no plateau in the extrapolated results and the difference
between the scalar and pseudoscalar is quite significant, it is difficult to
determine the quark mass from these plots.
We may need to exclude the data on the coarsest lattice from this analysis.

\begin{figure}[tbp]
\begin{center}
\subfigure{\mbox{\raisebox{1mm}{\includegraphics[width=110mm, bb=0 0 345 230]{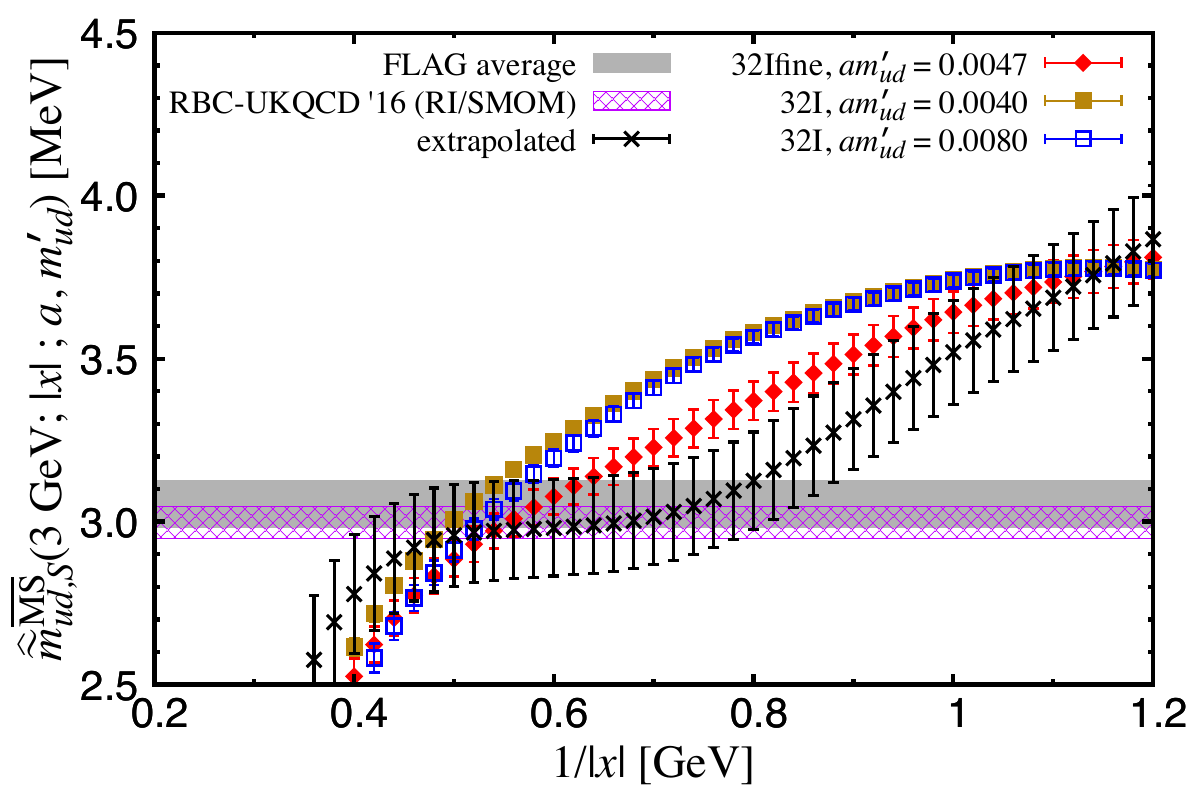}}}}
\subfigure{\mbox{\raisebox{1mm}{\includegraphics[width=110mm, bb=0 0 345 230]{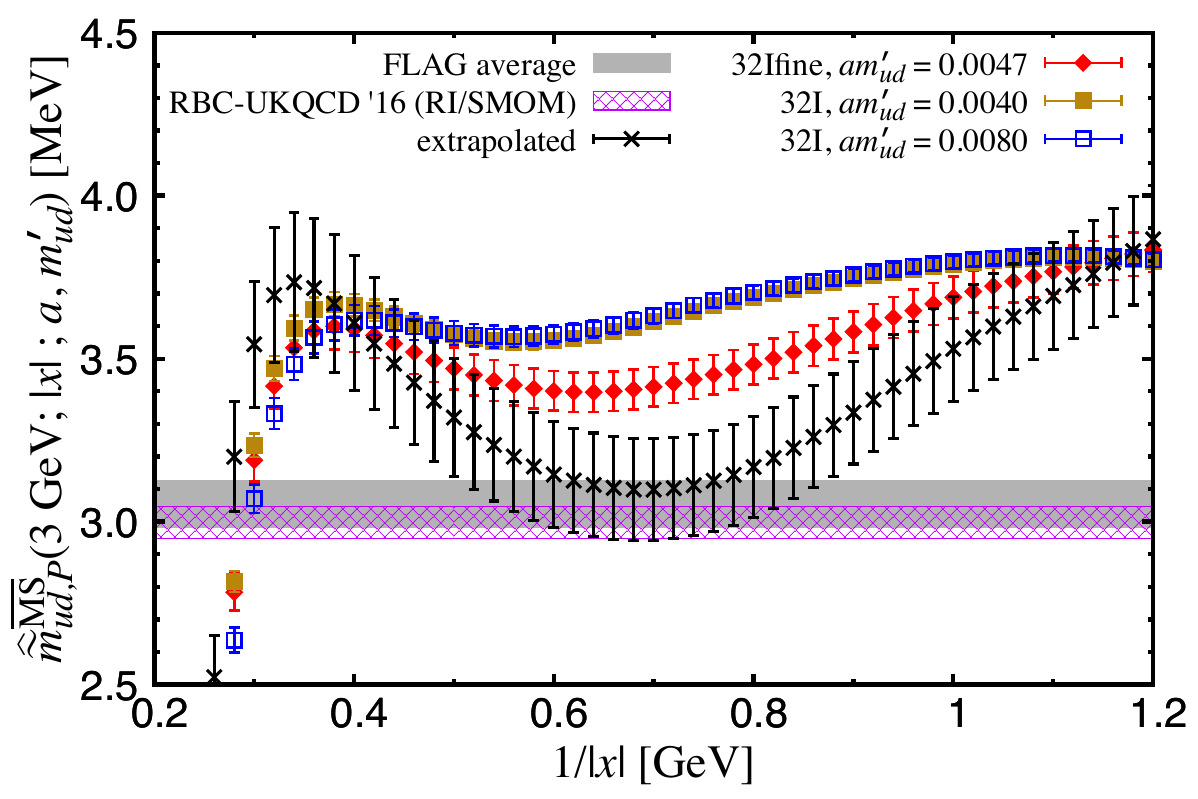}}}}
\caption{
Same as Figure~\ref{fig:mudMSb3GeV_ps} but the results only on the 32I and 32Ifine
ensembles are shown and used for the extrapolation.
}
\label{fig:mudMSb3GeV_ps_finer}
\end{center}
\end{figure}

Figure~\ref{fig:mudMSb3GeV_ps_finer} shows the results on the
finer two lattices (32I and 32Ifine).
The continuum limit is taken only with these lattice data, excluding the
result on the coarsest ensembles.
Since the number of free parameters in Eq.~\eqref{eq:extrapolation} and
the number of data points in this extrapolation are both three,
the extrapolation is not an actual $\chi^2$ fit but the free parameters can
still be determined, with propagated errors, by solving Eq.~\eqref{eq:extrapolation}.
While the $|x|$-dependence of the extrapolated result becomes milder
especially around $1/|x| \simeq0.8$~GeV, the statistical error is substantially
increased by discarding the data on the coarsest lattice.
In order to obtain a reasonable result with sufficiently small statistical error
from such an analysis, we need to introduce finer lattices.

Since it is currently not easy to introduce a finer lattice, we seek a more
economical analysis that enables to extract the quark mass from the
data we currently have.
Among the three sources of the $n$-dependence of
$\widetilde Z_{S/P}^{\rm\overline{MS}/lat}(\mu,1/a;an;m_{ud}')$ 
mentioned in Section~\ref{sec:mass_renorm}, we now focus on the
third one, the nonperturbative effects.
The nonperturbative effects on the scalar and pseudoscalar correlators
are known to be quite large compared to those on the vector and
axial-vector correlators in a model based on instantons because the scalar
and pseudoscalar channels
are directly affected by instantons~\cite{'tHooft:1976fv,Novikov:1981xi}.
The effect of a single instanton on these channels, which is the
most significant at short distances, is of the same magnitude but with
opposite sign~\cite{Shuryak:1993kg}.
Therefore, the na\"\i ve average of these two channels may be free from the
largest source of nonperturbative effects.
Therefore, we analyze
\begin{equation}
\widehat{\widetilde m}_q^{\rm\overline{MS}}({\rm 3~GeV};|x|;a,m_{ud}')
= \frac{m_q^{\rm bare}(1/a)}{\widehat{\widetilde Z}_{S+P}^{\rm\overline{MS}/lat}({\rm 3~GeV},1/a;|x|;m_{ud}')},
\label{eq:renorm_mass_ave}
\end{equation}
with the 
$O(4)$-symmetric renormalization factor
$\widehat{\widetilde Z}_{S+P}^{\rm\overline{MS}/lat}({\rm 3~GeV},1/a;|x|;m_{ud}')$
obtained from
Eqs.~\eqref{eq:ZsO4MSbar} and \eqref{eq:O4condition}
by substituting $S/P$ with $S+P$ and defining
$\widehat G_{S+P}^{\rm lat}(1/a;|x|)$ as the spherical average of
the average of the scalar and pseudoscalar correlators.

\begin{figure}[tbp]
\begin{center}
\subfigure{\mbox{\raisebox{1mm}{\includegraphics[width=110mm, bb=0 0 345 230]{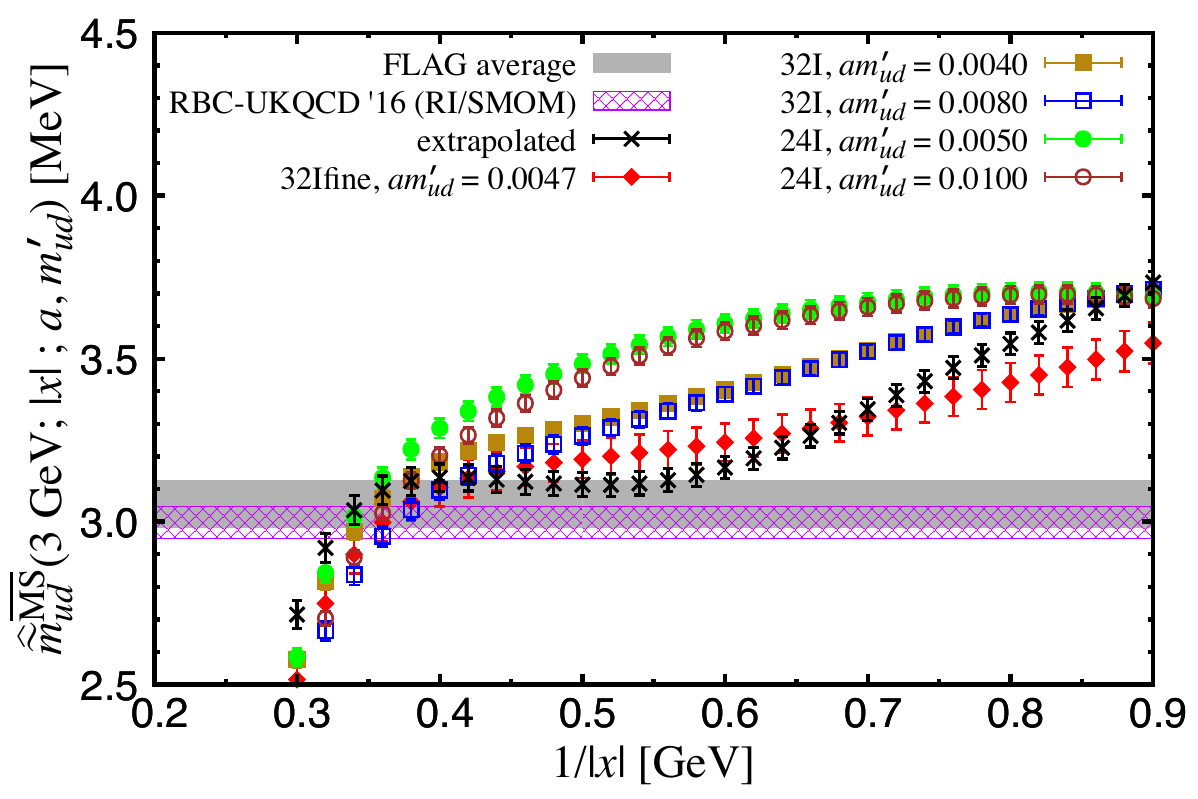}}}}
\subfigure{\mbox{\raisebox{1mm}{\includegraphics[width=110mm, bb=0 0 345 230]{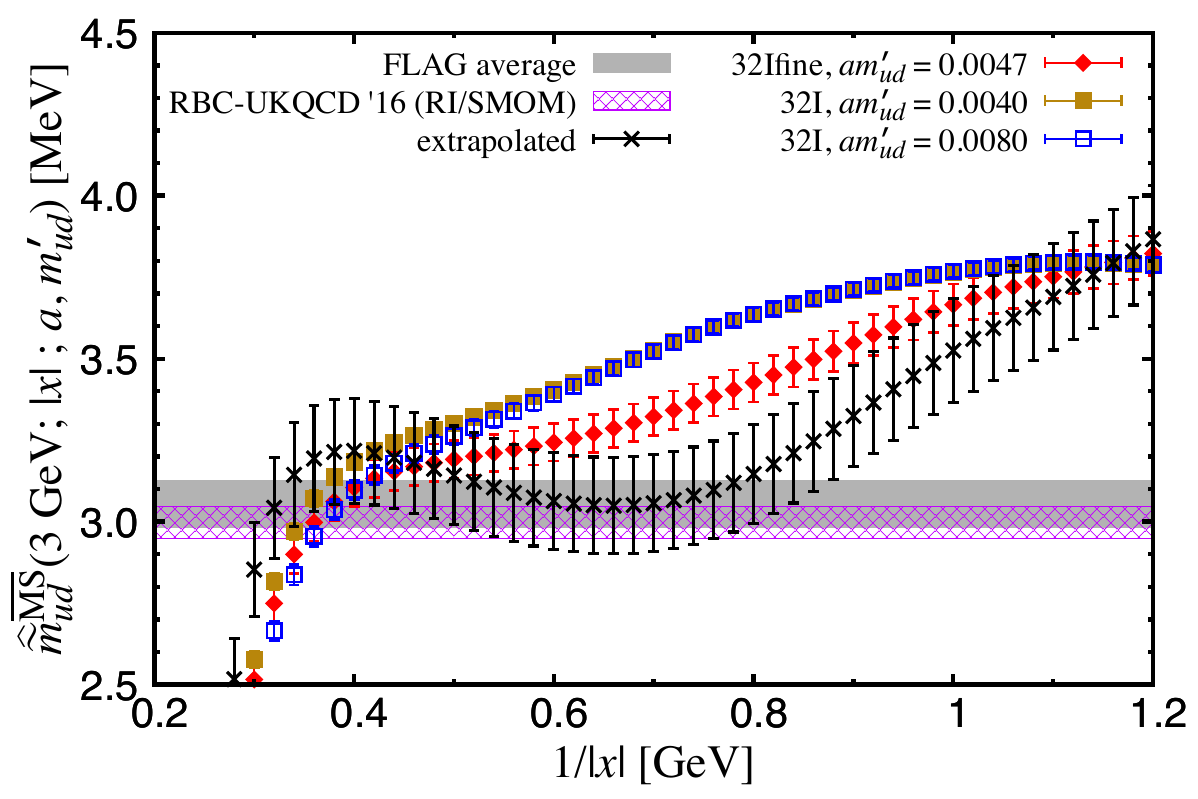}}}}
\caption{
Results for the spherical average
$\widehat{\widetilde m}_{ud}^{\rm\overline{MS}}(3~{\rm GeV};|x|;a,m_{ud}')$
calculated from the average of the scalar and pseudoscalar correlators.
The results for the extrapolation to the continuum and chiral
limits ($a\to0, m_{ud}'\to0$) performed using all the ensembles (upper panel)
and only the finer two lattices (lower panel) are also shown.
}
\label{fig:mudMSb3GeV}
\end{center}
\end{figure}

Figure~\ref{fig:mudMSb3GeV} shows the results for 
$\widehat{\widetilde m}_{ud}^{\rm\overline{MS}}({\rm 3~GeV};|x|;a,m_{ud}')$.
The continuum and chiral limits are taken using the fit function
\begin{equation}
\widehat{\widetilde m}_{ud}^{\rm\overline{MS}}(3~{\rm GeV};|x|;a,m_{ud}')
= \widehat{\widetilde m}_{ud}^{\rm\overline{MS}}(3~{\rm GeV};|x|)
+ C_{a}(|x|)a^2 + C_{m}(|x|)M_\pi(a,m_{ud}')^2,
\label{eq:extrapolation_ave}
\end{equation}
with the fit parameters
$\widehat{\widetilde m}_{ud}^{\rm\overline{MS}}(3~{\rm GeV};|x|)$,
$C_{a}(|x|)$ and $C_{m}(|x|)$.
The fit results with (upper panel) and without (lower panel) the data
on the coarsest lattice are shown.
We see a plateau of the extrapolated data in the interval 
${\rm 0.4~GeV}\lesssim1/|x|\lesssim0.6$~GeV in the upper panel and 
${\rm 0.4~GeV}\lesssim1/|x|\lesssim0.8$~GeV in the lower panel.
These facts agree with the instanton-based observation
that the nonperturbative effects on the average of the scalar and
pseudoscalar correlators are much smaller than those on the individual
channels.

\begin{figure}[tbp]
\begin{center}
\subfigure{\mbox{\raisebox{1mm}{\includegraphics[width=110mm, bb=0 0 345 230]{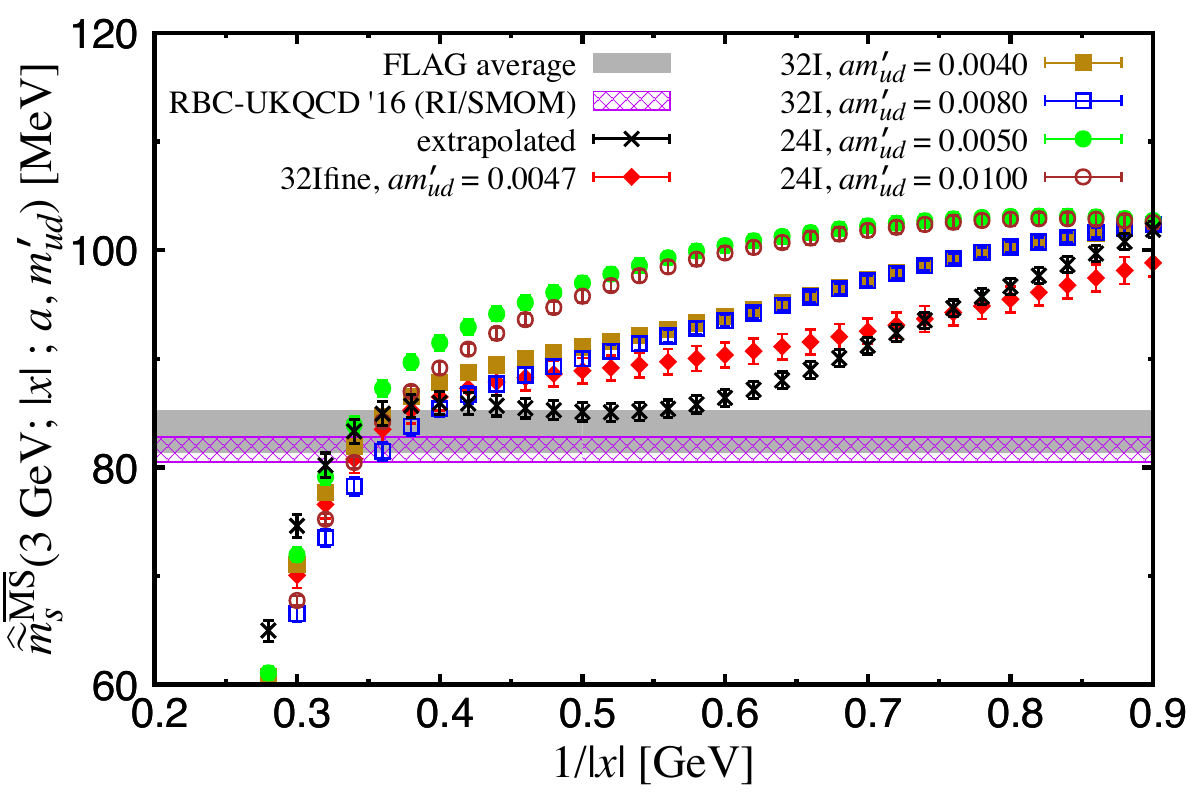}}}}
\caption{
Same as the upper panel of Figure~\ref{fig:mudMSb3GeV} but the result for
the strange quark mass.
}
\label{fig:msMSb3GeV}
\end{center}
\end{figure}

Figure~\ref{fig:msMSb3GeV} shows the result for the strange quark
mass defined in Eq.~\eqref{eq:renorm_mass_ave} with $q=s$,
where the same renormalization factors as for the light quark mass are used.
The continuum and chiral extrapolations are done by the fit function
\eqref{eq:extrapolation_ave} with the substitution
$\widehat{\widetilde m}_{ud}^{\rm\overline{MS}} \to \widehat{\widetilde m}_s^{\rm\overline{MS}}$.
A plateau is seen in the same region as in the result for the light quarks mass.

\begin{figure}[tbp]
\begin{center}
\subfigure{\mbox{\raisebox{1mm}{\includegraphics[width=110mm, bb=0 0 345 230]{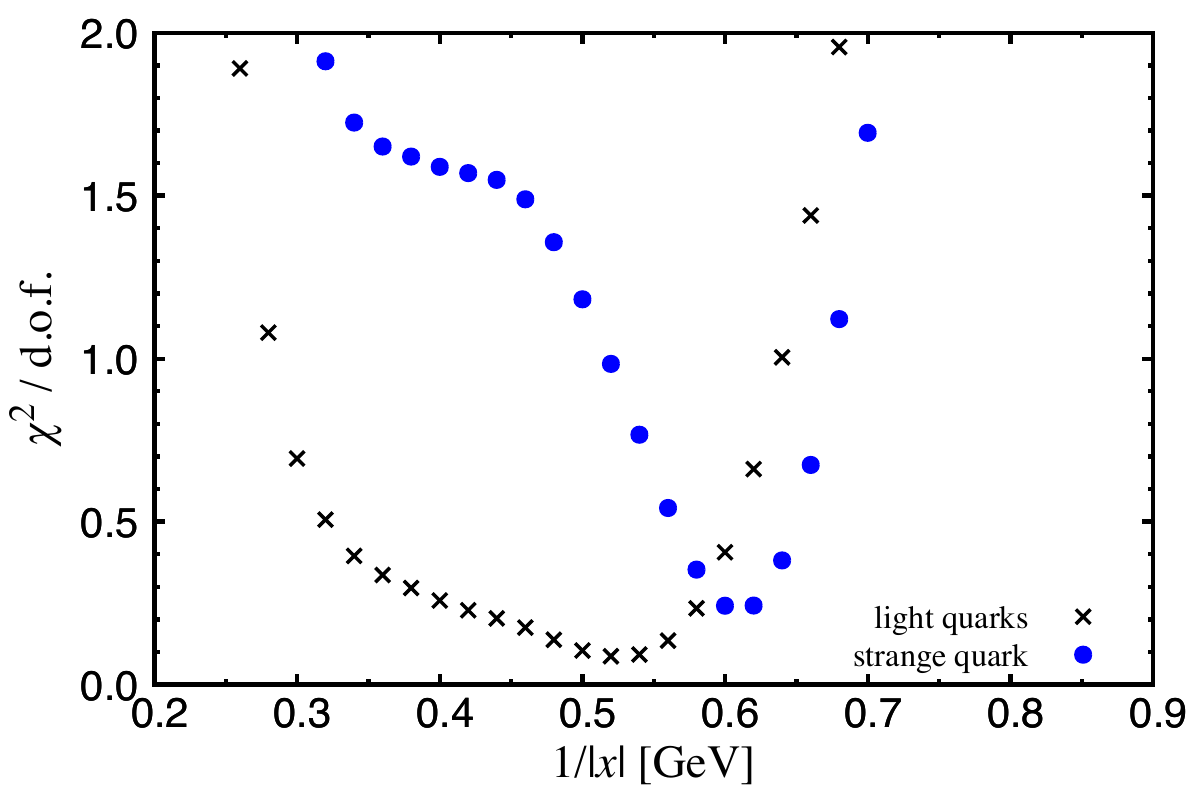}}}}
\caption{
The $\chi^2$/d.o.f. obtained through the fit in the
upper panel of Figure~\ref{fig:mudMSb3GeV} (crosses) and
in Figure~\ref{fig:msMSb3GeV} (circles).
}
\label{fig:fit_summary}
\end{center}
\end{figure}

Figure~\ref{fig:fit_summary} shows
the $\chi^2/$d.o.f.
obtained through the simultaneous fit both for the light (crosses) and
strange (circles) quark masses.
While the position-space renormalization factors calculated on each
ensemble are uncorrelated, the statistical errors for $Z_q$ and $m_q^{\rm bare,32I}$,
which are taken from Ref.~\cite{Blum:2014tka} and used in Eq.~\eqref{eq:def_mqbare}
to calculate unrenormalized quark mass $m_q^{\rm bare}(1/a)$ for each lattice cutoff,
are likely correlated.
We interpret the small values of $\chi^2/$d.o.f. shown in Figure 6
as resulting from our ignorance of such correlations.
Since $\chi^2$/d.o.f. at $1/|x|\simeq0.6$~GeV, which roughly 
corresponds to $|x|\simeq 3a$ on the coarsest ensemble,
is not too large,
we may conclude that the spherical average does not suffer
significantly from higher orders of $O(a)$ errors for $|x| \gtrsim 3a$.

\begin{figure}[tbp]
\begin{center}
\subfigure{\mbox{\raisebox{1mm}{\includegraphics[width=110mm, bb=0 0 345 230]{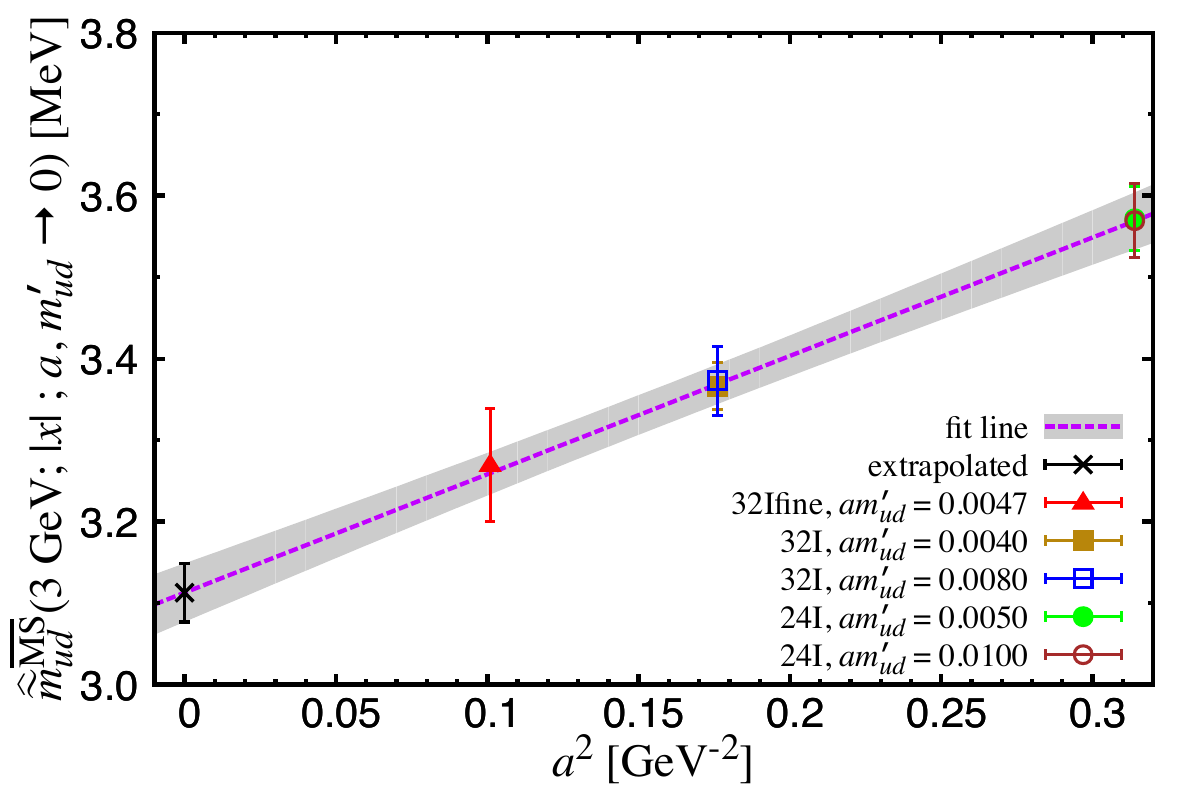}}}}
\subfigure{\mbox{\raisebox{1mm}{\includegraphics[width=110mm, bb=0 0 345 230]{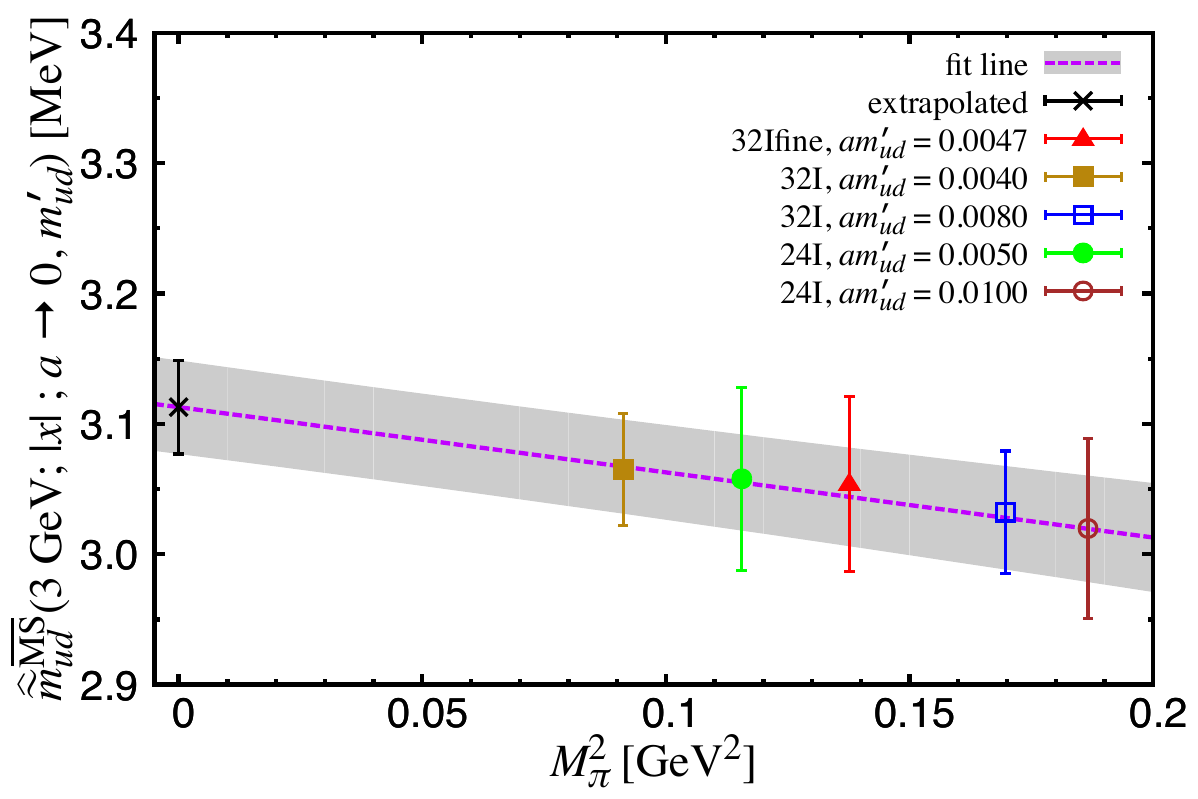}}}}
\caption{
Continuum and chiral extrapolations seen on the planes with $M_\pi = 0$
(upper panel) and $a=0$ (lower panel) at $1/|x| = 0.52$~GeV.
}
\label{fig:fit_summary2}
\end{center}
\end{figure}

To investigate the $a$- and $M_\pi$-dependences of 
$\widehat{\widetilde m}_{ud}^{\rm\overline{MS}}(3~{\rm GeV}; |x|;a,m_{ud}')$
more clearly, we visualize this extrapolation by analyzing the
renormalized mass at a specific distance $|x|$ in the chiral limit
\begin{align}
\widehat{\widetilde m}_{ud}^{\rm\overline{MS}}(3~{\rm GeV}; |x|;a,m_{ud}'\to0)
&\equiv
\widehat{\widetilde m}_{ud}^{\rm\overline{MS}}(3~{\rm GeV}; |x|;a,m_{ud}')
- C_mM_\pi(a,m_{ud}')^2
\notag\\&=
\widehat{\widetilde m}_{ud}^{\rm\overline{MS}}(3~{\rm GeV};|x|)
+ C_aa^2,
\end{align}
and in the continuum limit
\begin{align}
\widehat{\widetilde m}_{ud}^{\rm\overline{MS}}(3~{\rm GeV}; |x|;a\to0,m_{ud}')
&\equiv
\widehat{\widetilde m}_{ud}^{\rm\overline{MS}}(3~{\rm GeV}; |x|;a,m_{ud}')
- C_aa^2
\notag\\&=
\widehat{\widetilde m}_{ud}^{\rm\overline{MS}}(3~{\rm GeV};|x|)
+ C_mM_\pi(a,m_{ud}')^2,
\end{align}
with the parameters $\widehat{\widetilde m}_{ud}^{\rm\overline{MS}}(3~{\rm GeV};|x|)$,
$C_a(|x|)$ and $C_m(|x|)$ obtained through the simultaneous fit
Eq.~\eqref{eq:extrapolation_ave}.
Figure~\ref{fig:fit_summary2} shows the result for these values with
the lines of the extrapolation at $1/|x| = 0.52$~GeV.
The result indicates both of the $a$- and $M_\pi$-dependences are
treated well with their quadratic terms.

We use the result from the extrapolation including the data on the coarsest lattice
to determine the renormalized quark mass.
We estimate the central value of the quark mass at $1/|x| = 0.52$~GeV,
where $\chi^2/$d.o.f is minimum.
Our result is
\begin{equation}
m_{ud}^{\rm\overline{MS}}(3~{\rm GeV}) = 3.113(36)(52)(24)(70)\ \rm MeV.
\end{equation}
The first error is the statistical error.
The second error is the systematic error due to discretization effects,
which is estimated by increasing $1/|x|$ up to 0.60~GeV where a
deviation from the plateau is beginning.
The third error is the systematic uncertainty due to the truncation
of the perturbative calculation, which is estimated by varying the
parameter $\mu_x^*$ in the region
$\mu_x^{*,\rm opt}/\sqrt{2}\le\mu_x^*\le\sqrt{2} \mu_x^{*,\rm opt}$.
The forth error corresponds to the systematic error due to
the uncertainty of the strong coupling constant, which is estimated
by varying the scale of three-flavor QCD in the region
$315~{\rm MeV}\le\Lambda_{\rm QCD}^{\rm\overline{MS}}\le349$~MeV~
\cite{Tanabashi:2018oca}.
The result is compatible
with
our previous RBC/UKQCD result
$m_{ud}^{\rm\overline{MS}}(3~{\rm GeV}) = 2.997(49)$~MeV~\cite{Blum:2014tka}
obtained through the RI/SMOM scheme
using the same lattice ensembles
and with the FLAG average
$m_{ud}^{\rm\overline{MS}}(3~{\rm GeV}) = 3.054(72)$~MeV~\cite{Aoki:2016frl}
of many works done through various renormalization
procedures including the RI/(S)MOM and the Schr\"odinger
functional~\cite{Luscher:1992an} methods using $2+1$-flavor ensembles.
Applying the same procedure, we obtain the strange quark mass
\begin{equation}
m_s^{\rm\overline{MS}}(3~{\rm GeV}) = 85.07(84)(1.33)(66)(1.92)\ \rm MeV,
\end{equation}
which is also compatible
with
the FLAG average 
$m_s^{\rm\overline{MS}}(3~{\rm GeV}) = 83.3(1.9)$~MeV~\cite{Aoki:2016frl}
and
our previous RBC/UKQCD result 
$m_s^{\rm\overline{MS}}(3~{\rm GeV}) = 81.64(1.17)$~MeV~\cite{Blum:2014tka}
determined through the RI/SMOM scheme
using the same ensembles.

\tabcolsep = 6pt
\begin{table}[tbp]
\caption{
Values of $\hat{\widetilde Z}_m^{\rm\overline{MS}/lat}({\rm3~GeV},1/a;|x|;m_{ud}') = 1/\hat{\widetilde Z}_{P+S}^{\rm\overline{MS}/lat}({\rm3~GeV},1/a;|x|;m_{ud}')$ calculated on each ensemble at $1/|x| = 0.44$~GeV, 0.52~GeV and 0.60 GeV.
The given uncertainties (from left to right) are statistical, one from truncation of the perturbative calculation and one from $\Lambda_{\rm QCD}^{\rm\overline{MS}}$.
}
\label{tab:Zm}
\begin{center}
\begin{tabular}{|cc|ccc|}
\hline
\multirow{2}{*}{Ensemble set} & \multirow{2}{*}{$m_{ud}'$}
& \multicolumn{3}{c|}{$1/|x|$ [GeV]} \\ \cline{3-5}
& & 0.44 & 0.52 & 0.60 \\
\hline
24I & 0.0050 & 1.495(6)(25)(51) & 1.553(5)(12)(35) & 1.595(4)(7)(26) \\
    & 0.0100 & 1.467(2)(24)(50) & 1.537(2)(12)(35) & 1.584(2)(7)(26) \\
\hline
32I & 0.0040 & 1.475(7)(25)(51) & 1.511(5)(12)(34) & 1.549(4)(7)(25) \\
    & 0.0080 & 1.446(11)(24)(50) & 1.496(9)(12)(34) & 1.542(7)(7)(25) \\
\hline
32Ifine & 0.0047 & 1.456(9)(24)(50) & 1.478(6)(11)(33) & 1.498(4)(7)(24) \\
\hline
\end{tabular}
\end{center}
\end{table}

In Table~\ref{tab:Zm},
we summarize the quark mass renormalization factors calculated on
each ensemble and at three values of $|x|$.
The three errors shown (from left to right) correspond to the statistical error,
the systematic error due to the truncation of the perturbative calculation and
due to the uncertainty of the QCD scale $\Lambda_{\rm QCD}^{\rm\overline{MS}}$,
in order.
The latter two uncertainties, which are associated with perturbation theory,
are more significant at longer distances as one can easily expect.
As we have mentioned throughout the paper, we need to fix $|x|$ when we 
renormalize a quantity and take its continuum and chiral limits since 
the structure of $a$- and $m_{ud}'$-dependences depends on $|x|$.


\section{Conclusion}
\label{sec:conclusion}

We have proposed a spherical averaging technique for position-space
renormalization to reduce discretization errors and enhance the
renormalization window.  
This technique has the further important advantage that it allows
renormalized quantities to be defined at any fixed physical distance.  
This allows a direct matching between renormalized quantities defined
on ensembles with different lattice spacings and a continuum limit to
be easily taken for position-space renormalized quantities at a fixed,
physical renormalization scale.

The technique is applied to the quark mass renormalization using
the scalar and pseudoscalar correlators in position space.
We investigate the $|x|$-dependence of the renormalized quark
mass in the continuum limit and find a plateau even when
the $\rm\overline{MS}$ renormalized quark mass at finite $a$
still depends slightly on the distance $|x|$ at which the intermediate
position-space scheme is applied.
The investigation of the $\chi^2$/d.o.f. found for the continuum and
chiral extrapolations implies 
that the $a$-dependence of sphere-averaged correlator is mostly
$O(a^2)$ in the region $|x|\gtrsim3a$.
The renormalized quark mass obtained through this renormalization
procedure agrees with the FLAG average and our previous RBC/UKQCD
result obtained by using the RI/SMOM renormalization scheme
on the same lattice ensembles.

The averaging approach proposed here is one of many possible schemes
that can be devised which involve a smearing or averaging over lattice
points in position space.  
However, the scheme proposed here may be of particular value because
it involves two quite sharply defined scales: a long-distance scale $|x|$,
the radius of the sphere over which we average, and a short-distance scale,
the lattice spacing $a$ which describes the thickness of the spherical shell
of points which are averaged.  
The multi-linear interpolation method which is employed might be viewed as
among the simplest prescriptions for creating this average.  
Having two such distinct scales may improve the continuum limit of the
quantities renormalized using this method.

Since this X-space scheme is gauge invariant and
free from contact terms, it prevents the mixing with irrelevant
operators, which can be a serious complication for gauge-noninvariant
schemes such as the RI/MOM scheme.  
Therefore, this position-space renormalization is especially well-suited
for the four-quark operators in the $\Delta S=1$ weak Hamiltonian
where it can be imposed at the relatively long distances needed to
define three-flavor operators --- distances much longer than the
Compton wavelength of the charm quark.
At such distances (or at the corresponding energies below 1 GeV), the
RI/MOM scheme is plagued by gauge noise and the usually justified
neglect in the RI/MOM scheme of additional dimension-six operators
constructed from a product of quark bilinears and gluon fields is likely
to be a poor approximation.  
In fact, one of the motivations for this X-space method is to allow the
Wilson coefficients of the three-flavor $\Delta S=1$ weak Hamiltonian to
be determined non-perturbatively in terms of the more accurate,
perturbatively-determined Wilson coefficients of the corresponding
four-flavor theory.
 
Further technique must be developed before such a complete three-to-four
flavor matching is possible.  
Some renormalization conditions can be imposed on the position-space
two-point functions of four-quark operators renormalized in analogy with
the sphere average of the two-point functions of scalar or pseudo-scalar
currents presented in this paper.  
However, such in a position-space scheme, we will also need to constrain
other Green's functions such as three-point functions of a four-quark
operator and two two-quark operators to uniquely define a position-space scheme. 
The two-point functions of $N$ mixing operators, will form a real symmetric matrix
and allow at most $N(N+1)/2$ renormalization conditions to be imposed.  
These will be insufficient to determine the needed $N\times N$ renormalization
matrix. 
An extension of the spherical averaging procedure to such three-point functions
is the next step with is being developed.


\begin{acknowledgments}
The authors thank their RBC and UKQCD colleagues for many useful
discussions especially Peter~Boyle and Robert Mawhinney.
This work is supported in part by the US DOE grant \#DE-SC0011941.
\end{acknowledgments}

\appendix

\section{Irregular $a$-dependence of spherical average}
\label{app:side_effect}

Since the interpolation Eq.~\eqref{eq:quadrilin_interpolate} contains
$n_\mu = \lfloor x_\mu/a\rfloor$, which is a discontinuous function of $x/a$,
some irregularity could occur and the continuum extrapolation
with only a term proportional to $a^2$ may not be accurate.
In this Appendix, we discuss the significance of such irregular
$a$-dependence of the spherical average.

Let us begin with the case of one dimension, in which the
interpolated value $\bar f(x)$ of a quantity $f(x)$ defined in
Eq.~\eqref{eq:eq_wave1dim} can be written as
\begin{equation}
\bar f_a(x) = F(x) + c_a(x)a^2 + {x^2F''(x)\over2}\left(\frac{a}{x}(n+1)-1\right)\left(1-\frac{a}{x}n\right) + O(a^3).
\label{eq:1dim_nform}
\end{equation}
Here, $F''(x)$ stands for the second derivative of $F(x)$ and $c_a(x)$ comes from
$c_{a,n}$ and $c_{a,n+1}$ in Eq.~\eqref{eq:lattice_error}, which are the discretization
errors in the values $f_{a,n}$ and $f_{a,n+1}$ evaluated on the lattice.
We omit the possible logarithmic $a$-dependence of $F(x)$ and $F''(x)$ for simplicity.

\begin{figure}[tbp]
\begin{center}
\subfigure{\mbox{\raisebox{1mm}{\includegraphics[width=110mm, bb=0 0 345 230]{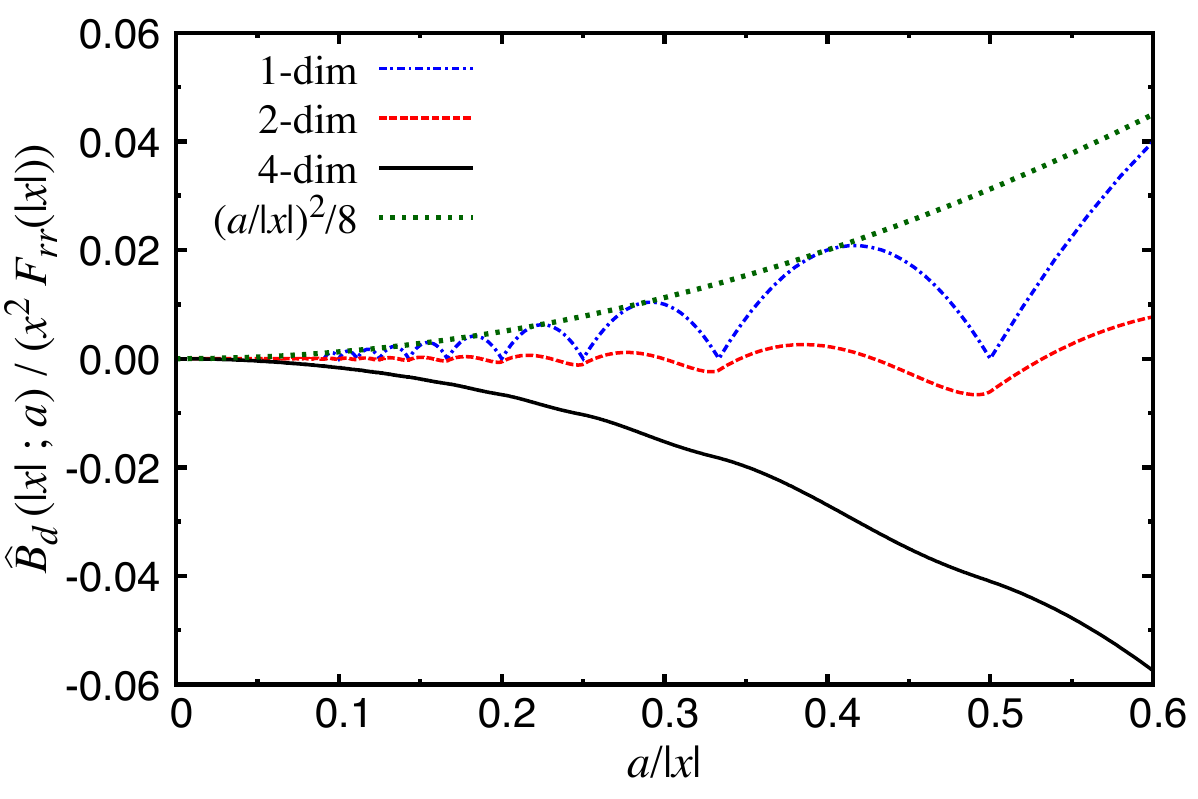}}}}
\caption{
$B_1(x;a)/(x^2F''(x))$, $\widehat B_2(|x|;a)/(x^2F_{rr}(|x|))$ and $\widehat B_4(|x|;a)/(x^2F_{rr}(|x|))$
plotted as functions of $a/|x|$.
The curve of $(a/|x|)^2/8$ is also plotted for comparison.
}
\label{fig:third_term_2dim}
\end{center}
\end{figure}

While the both of the second and third terms in Eq.~\eqref{eq:1dim_nform}
are expected to contain complicated $a$-dependence, the third term
\begin{equation}
B_1(x;a) = {x^2F''(x)\over2}\left(\frac{a}{x}(n+1)-1\right)\left(1-\frac{a}{x}n\right),
\label{eq:F1}
\end{equation}
can be explicitly analyzed and therefore is discussed first.
Since the value of $n=\lfloor x/a\rfloor$ jumps where $x/a$ is an integer,
the $a$-dependence of $I_1(x;a)/(x^2F''(x))$ given by Eq.~\eqref{eq:F1}
is drawn (dashed curve) in Figure~\ref{fig:third_term_2dim}.
Thus, the continuum extrapolation with a few data points with an
assumption of simple $a^2$ discretization error could be inaccurate.
While such ambiguity is expected to be less than 1\% of
$x^2F''(x)$ in the case of one dimension, one could imagine that in
higher dimensions the spherical average further softens such irregular
dependence on $a$ and makes the continuum extrapolation more accurate.

The interpolation in $d$ dimensions can be written as
\begin{equation}
\bar f_a(x) = F(x) + c_a(x)a^2 + B_d(x;a) + O(a^3),
\label{eq:ddim_nform}
\end{equation}
where
\begin{equation}
B_d(x;a) = {x^2\over2}
\sum_{\mu=1}^d
F_{\mu\mu}(x)\left(\frac{a}{|x|}(n_\mu+1)-\frac{x_\mu}{|x|}\right)\left(\frac{x_\mu}{|x|}-\frac{a}{|x|}n_\mu\right),
\end{equation}
and $F_{\mu\mu}(x)$ is the second derivative of $F(x)$ with respect to $x_\mu$.
Averaging over the sphere by the integral given in Eq.~\eqref{eq:sphe_ave2} for
two dimensions or in Eq.~\eqref{eq:sphe_ave4} for four dimensions,
this term becomes
\begin{align}
\widehat B_d(|x|;a) &=
C_d\,x^2
\int_0^{\pi/2}\td\theta
\sin^{d-2}\theta
\left(
F_{rr}(|x|)\cos^2\theta+{F_r(|x|)\over |x|}\sin^2\theta
\right)
\notag\\*
&\hspace{16mm}
\times\left(
\frac{a}{|x|}\left(
\left\lfloor
{|x|\over a}\cos\theta
\right\rfloor+1
\right)-\cos\theta
\right)
\left(
\cos\theta-\frac{a}{|x|}
\left\lfloor
{|x|\over a}\cos\theta
\right\rfloor
\right),
\label{eq:ext_term_phave4d}
\end{align}
where
\begin{equation}
C_2 = \frac{2}{\pi},\ \ 
C_4 = \frac{8}{\pi},
\end{equation}
and $F_r(|x|)$ and $F_{rr}(|x|)$ respectively stand for the first and second
derivatives of $F(|x|)$ with respect to $|x|$.

\begin{figure}[tbp]
\begin{center}
\subfigure{\mbox{\raisebox{1mm}{\includegraphics[width=110mm, bb=0 0 345 230]{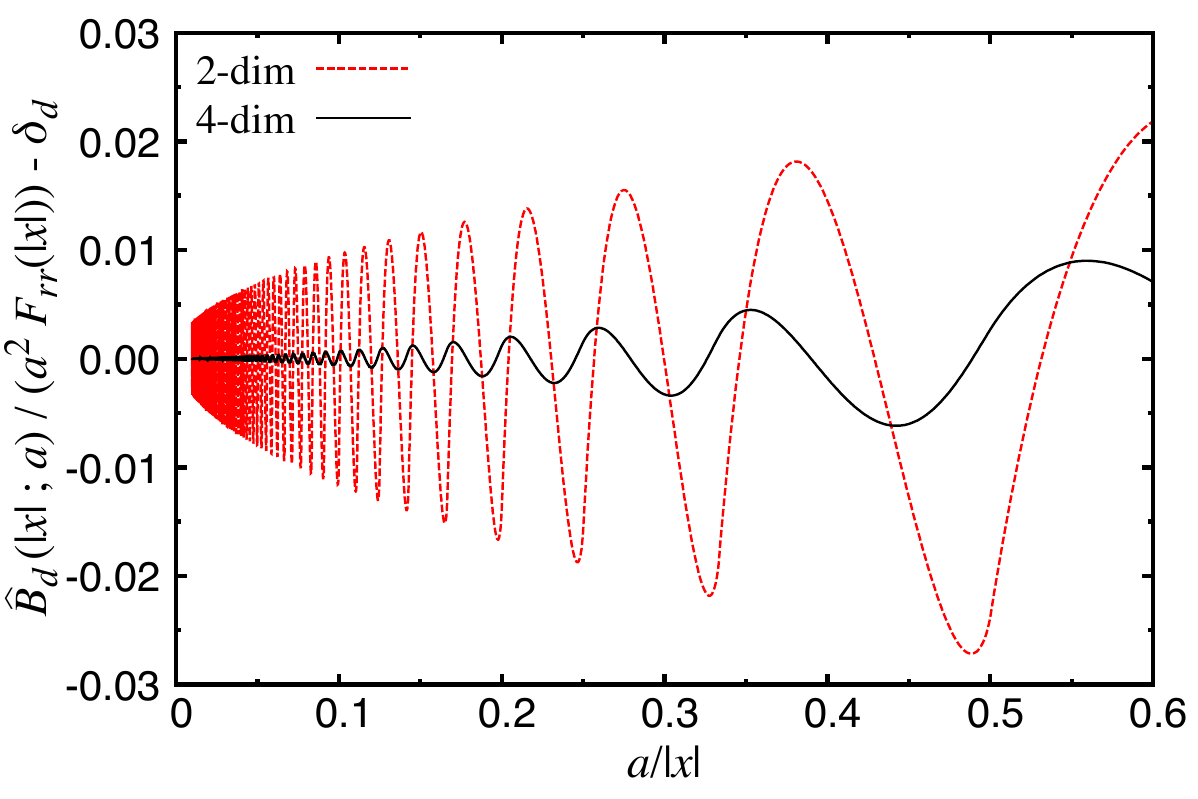}}}}
\caption{
$B_2(|x|;a)/(a^2F_{rr}(|x|))$ and $B_4(|x|;a)/(a^2F_{rr}(r))$ shifted by $\delta_d$,
the mid point of the oscillation.
Note by dividing by $a^2$ instead of by $|x|^2$ as is done in
Figure~\ref{fig:third_term_2dim}, we are plotting the correction
relative to the regular $a^2$ error.
}
\label{fig:third_term_scale}
\end{center}
\end{figure}

In order to describe the $a$ dependence implied by
Eq.~\eqref{eq:ext_term_phave4d}, we need to assume a relation
between the $F_{rr}(|x|)$ and $F_r(|x|)/|x|$ terms found in the first line
of that equation.  For the purposes of illustration we will assume that
$F(|x|)$, which in this application is expected to be a slowly varying
function of $|x|$, behaves as $\ln(|x|)$ so we can replace  $F_r(|x|)$
by the $|x|F_{rr}(|x|)$.
In this case, Eq.~\eqref{eq:ext_term_phave4d} becomes
\begin{align}
\widehat B_d(|x|;a) &\simeq
C_dx^2F_{rr}(|x|)
\int_0^{\pi/2}\td\theta
\sin^{d-2}\theta
\left(
2\cos^2\theta-1
\right)
\notag\\*
&\hspace{16mm}
\times\left(
\frac{a}{|x|}\left(
\left\lfloor
{|x|\over a}\cos\theta
\right\rfloor+1
\right)-\cos\theta
\right)
\left(
\cos\theta-\frac{a}{|x|}
\left\lfloor
{|x|\over a}\cos\theta
\right\rfloor
\right).
\end{align}
Figure~\ref{fig:third_term_2dim} also shows the results for the spherical average
$\widehat B_d(|x|;a)$ normalized by $x^2F_{rr}(|x|)$ in two and four dimensions.
The magnitude of the irregular $a$-dependence in two dimensions is smaller than
that in one dimension and therefore the continuum extrapolation with a regular $a^2$
term is expected to be more accurate.
The irregular $a$-dependence of the spherical average in four dimensions is
even smaller than that in two dimensions in the sense explained below.
The significance of the irregular term in
\begin{equation}
\frac{\widehat B_d(|x|;a)}{x^2F_{rr}(|x|)} = \delta_d\cdot(a/|x|)^2 + (\mbox{irregular oscillation}) + O\left((a/|x|)^3\right),
\end{equation}
can be investigated by
\begin{equation}
\frac{\widehat B_d(|x|;a)}{a^2F_{rr}(|x|)} - \delta_d,
\end{equation}
which is plotted in Figure~\ref{fig:third_term_scale} for two and four dimensions.
Here, we find $\delta_2 \simeq 0.00047$ and $\delta_4 \simeq 0.16667$.
As the figure indicates, the irregular oscillation in four dimensions is even
smaller than that in two dimensions.
It means the continuum limit can be safely taken with a regular $a^2$ term.

\begin{figure}[tbp]
\begin{center}
\subfigure{\mbox{\raisebox{1mm}{\includegraphics[width=110mm, bb=0 0 345 230]{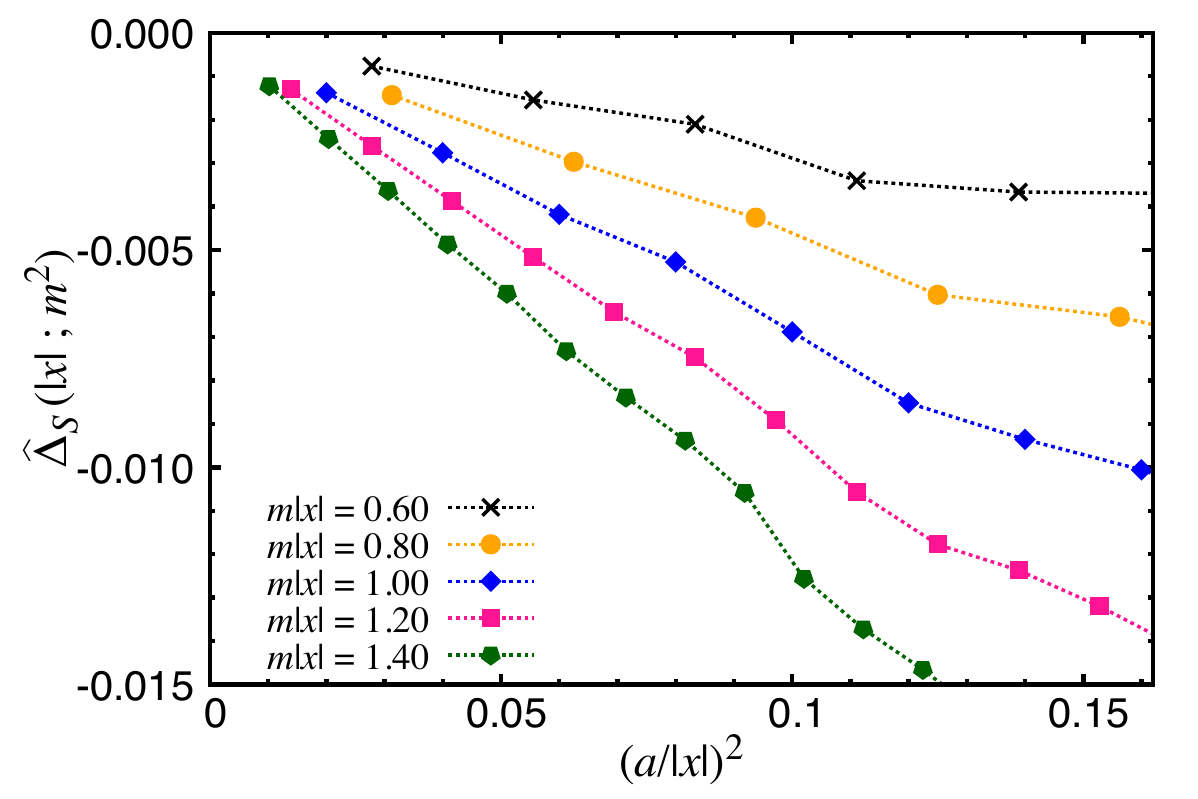}}}}
\caption{
$\widehat\Delta_S(|x|;m^2)$ versus $(a/|x|)^2$ for several values of $m|x|$.
}
\label{fig:Delta_S}
\end{center}
\end{figure}

Thus, we conclude that the irregular $a$-dependence associated with
the third term of Eq.~\eqref{eq:1dim_nform} or \eqref{eq:ddim_nform} is
negligible.
We proceed to discuss the other source of the irregular $a$-dependence
associated with $c_a(x)$ in Eq.~\eqref{eq:1dim_nform} or \eqref{eq:ddim_nform},
which can be written as
\begin{equation}
c_a(x) = a^{-4}\sum_{i,j,k,l=0}^1
\Delta_{1,i}\Delta_{2,j}\Delta_{3,k}\Delta_{4,l}\,
c_{a,n+i\hat1+j\hat2+k\hat3+l\hat4},
\end{equation}
in the case of four dimensions. 
Here, $c_n$ for the scalar or pseudoscalar correlator may
be roughly approximated to a dispersion integral of the difference
between the lattice and continuum propagators of a scalar field
\begin{align}
c_{a,n}a^2 &= 
\int_0^\infty\td s\, \rho(s)\delta D_F(an;s),\\
\delta D_F(an;m^2) &= D_F^{\rm lat}(an;m^2) - D_F^{\rm cont}(an;m^2),
\\
D_F^{\rm lat}(an;m^2) &= \int_{-\pi/a}^{\pi/a}\frac{\td^4q}{(2\pi)^4}\e^{\img aqn}\frac{1}{4a^{-2}\sum_\mu\sin^2{aq_\mu\over2}+m^2},
\\
D_F^{\rm cont}(x;m^2) &= \int_{-\infty}^\infty\frac{\td^4q}{(2\pi)^4}\e^{\img qx}\frac{1}{q^2+m^2}
= \frac{m}{4\pi^2}\frac{K_1(m|x|)}{|x|},
\end{align}
where $\rho(s)$ is the corresponding spectral function and $K_1(z)$
is the modified Bessel function of the second kind.
To quantify the significance of $c_n$, we analyze the spherical average
$\widehat\Delta_S(|x|;m^2)$ of
\begin{equation}
\Delta_S(an;m^2) = \frac{\delta D_F(an;m^2)}{D_F^{\rm cont}(x;m^2)\big|_{x=an}}.
\end{equation}
The result is shown in Figure~\ref{fig:Delta_S}, which indicates that
the discretization error is mostly proportional to $a^2$ for $|x| \gtrsim 3a$
and that the continuum extrapolation using lattice data at $|x| \simeq 3a,4a$
and $5a$ is likely accurate within the $O(0.1\%)$ level.

\begin{figure}[tbp]
\begin{center}
\subfigure{\mbox{\raisebox{1mm}{\includegraphics[width=110mm, bb=0 0 345 230]{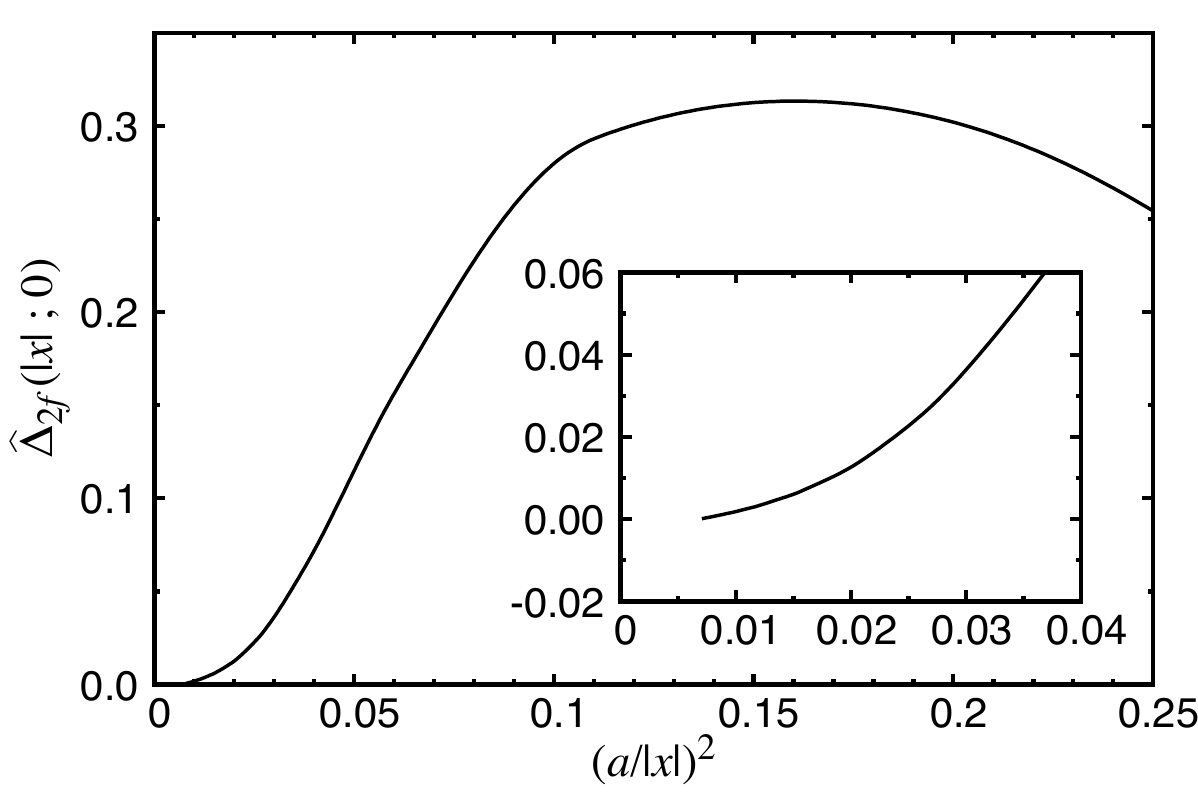}}}}
\subfigure{\mbox{\raisebox{1mm}{\includegraphics[width=110mm, bb=0 0 345 230]{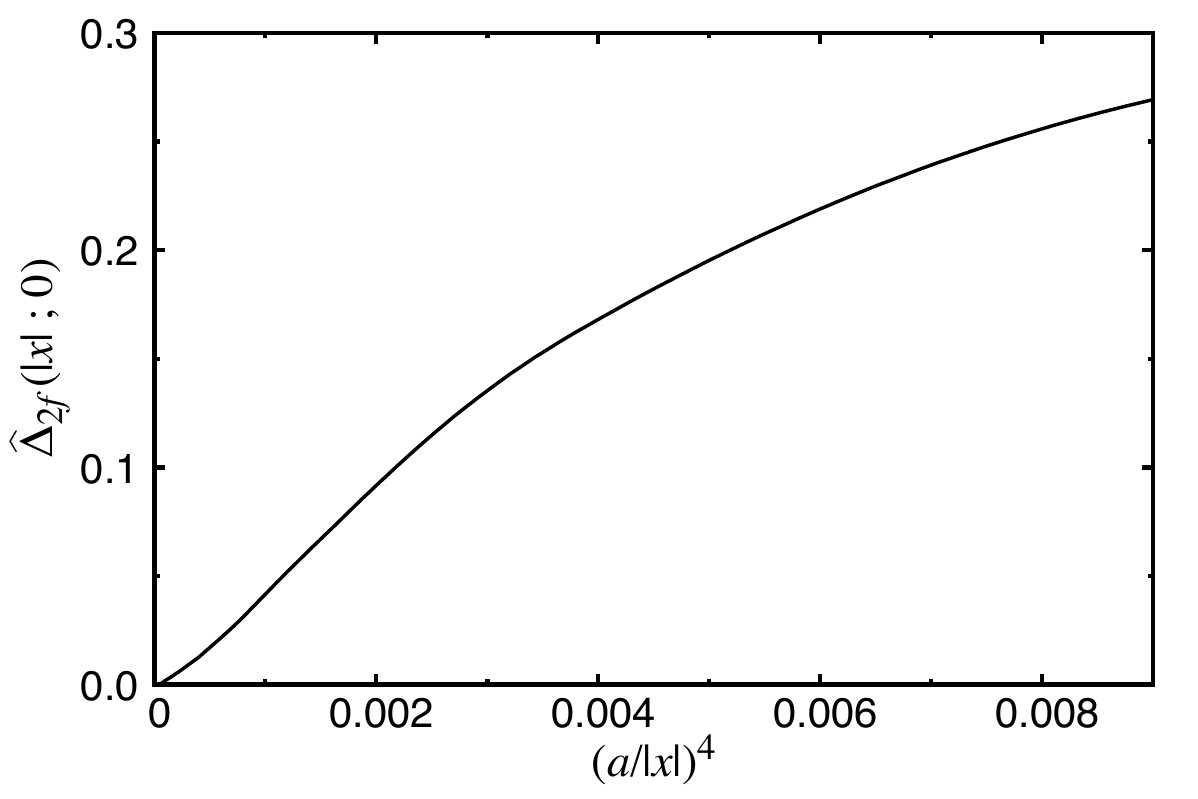}}}}
\caption{
$\widehat\Delta_{2f}(|x|;0)$ as a function of $(a/|x|)^2$ (upper panel) and
$(a/|x|)^4$ (lower panel).
}
\label{fig:Delta_2f}
\end{center}
\end{figure}

While the above analysis using the bosonic propagator may be valid in QCD
at long distances, the discretization error of Green's functions in the perturbative
regime may need to be discussed in terms of fermionic propagators.
We analyze the spherical average $\hat\Delta_{2f}(|x|;0)$ of 
\begin{align}
\Delta_{2f}(an;0) &= \lim_{m\to0}
\frac{G_S^{\rm lat,free}(an;m)-G_S^{\rm cont,free}(x;m)}{G_S^{\rm cont,free}(x;m)}
\Bigg|_{x=an},
\\
G_S^{\rm cont,free}(x;m)
&= \frac{3}{\pi^4x^6},
\\
G_S^{\rm lat,free}(x;m)
&= \Tr\left[
S_F^{\rm lat,free}(x;m)S_F^{\rm lat,free}(-x;m)
\right],
\end{align}
where the lattice propagator $S_F^{\rm lat,free}(an;m)$ in free field
theory at small input mass $am\lesssim0.1$ is quite sensitive to finite
volume since the physical length scale in the deconfinement phase
is associated with the input quark mass, not the pion mass.
Here, we calculate it as the Fourier transform of the corresponding
momentum-space propagator of domain-wall fermions
\cite{Narayanan:1992wx,Shamir:1993zy}.
The calculation on a $200^4$ lattice at $am\sim0.005$ still
suffers from significant finite volume effects, which can mostly be
estimated as the effect of the fermions wrapping around the volume
in the continuum theory.
See \cite{Tomii:2016xiv} for more detail.
Figure~\ref{fig:Delta_2f} shows the result for the spherical average
$\hat\Delta_{2f}(|x|;0)$ plotted as a function of $(a/|x|)^2$ (upper panel)
and $(a/|x|)^4$ (lower panel).Although the $a$-dependence at small
values of $a$ appears to be proportional to $a^4$, the coefficient
is somehow large $\sim 50(a/|x|)^4$.
We might therefore need to be careful when we take the
continuum limit. 
Of course, nonperturbative interactions could reduce the magnitude
of the $O(a^4)$ term to a size closer to that found above using the
bosonic propagator.

\bibliography{NPR_Zm.bbl}
\end{document}